\documentclass[lettersize,journal]{IEEEtran}

\usepackage{cite}
\usepackage{amsmath,amssymb,amsfonts}
\usepackage{graphicx}
\usepackage{textcomp}
\usepackage{xcolor}
\usepackage{authblk}

\usepackage{booktabs}
\usepackage{epsfig}
\usepackage{latexsym}
\usepackage{multirow}
\usepackage{stfloats}
\usepackage{epstopdf}
\usepackage{color}  
\usepackage{tabularx} 
\usepackage{enumerate}
\usepackage{array}
\graphicspath{{./Figures/}}
\usepackage{color}
\usepackage{bbm}
\usepackage{caption}
\usepackage{bm}
\usepackage[tight,footnotesize]{subfigure}
\usepackage{balance}
\usepackage{mathrsfs}
\usepackage{verbatim}
\allowdisplaybreaks[4]
\usepackage{dsfont}
\usepackage{verbatim}
\usepackage{tikz}
\usepackage{setspace}
\usepackage{diagbox}
\usepackage[framemethod=tikz]{mdframed}
\usepackage{multicol}
\usepackage{environ}
\usepackage{tikz}
\usepackage{stfloats}
\usepackage{algpseudocode}
\usepackage{graphics}
\usepackage{epsfig}
\usepackage{amsthm}
\usepackage{authblk}
\usepackage{enumitem}
\usepackage{makecell}

\ifCLASSOPTIONcompsoc
\usepackage[caption=false, font=normalsize, labelfont=sf, textfont=sf]{subfig}
\else
\usepackage[caption=false, font=footnotesize]{subfig}

\usepackage{bm}
\def\BibTeX{{\rm B\kern-.05em{\sc i\kern-.025em b}\kern-.08em
    T\kern-.1667em\lower.7ex\hbox{E}\kern-.125emX}}
\begin{document}

\title{Movable Antennas: Channel Measurement, Modeling, and Performance Evaluation}

\author{Yiqin~Wang,
Heyin~Shen,
Chong~Han,~\IEEEmembership{Senior~Member,~IEEE},
and~Meixia~Tao,~\IEEEmembership{Fellow,~IEEE}%
\thanks{Yiqin Wang and Heyin Shen are with Terahertz Wireless Communications (TWC) Laboratory, Shanghai Jiao Tong University, China (Email: \{wangyiqin, heyin.shen\}@sjtu.edu.cn).}%
\thanks{Chong Han is with Terahertz Wireless Communications (TWC) Laboratory, also with Department of Electronic Engineering and the Cooperative Medianet Innovation Center (CMIC), Shanghai Jiao Tong University, China (Email: chong.han@sjtu.edu.cn).}%
\thanks{Meixia Tao is with the Department of Electronic Engineering, Shanghai Jiao Tong University, Shanghai, China (E-mail: mxtao@sjtu.edu.cn).}}

\maketitle
\begin{abstract}

Since decades ago, multi-antenna has become a key enabling technology in the evolution of wireless communication systems. In contrast to conventional multi-antenna systems that contain antennas at fixed positions, position-flexible antenna systems have been proposed to fully utilize the spatial variation of wireless channels. In this paper, movable antenna (MA) systems are analyzed from channel measurement, modeling, position optimization to performance evaluation.
First, a broadband channel measurement system with physical MAs is developed, for which the extremely high movable resolution reaches 0.02~mm. A practical two-ray model is constructed based on the channel measurement for a two-dimensional movable antenna system across $32\times32$ planar port positions at 300~GHz.
In light of the measurement results, spatial-correlated channel models for the two-dimensional MA system are proposed, 
which are statistically parameterized by the covariance matrix of measured channels.
Finally, the signal-to-interference-and-noise ratio (SINR)-maximized position selection algorithm is proposed, which achieves 99\% of the optimal performance. The performance of different MA systems in terms of spectral efficiency are evaluated and compared for both planar and linear MA systems. Extensive results demonstrate the advantage of MAs over fixed-position antennas in coping with the multi-path fading and improving the spectral efficiency by 10\% in a 300~GHz measured channel.


\end{abstract}

\section{Introduction}

\par Over the past few decades, multi-antenna has become a key enabling technology in the evolution of wireless communication systems. From multiple-input multiple-output (MIMO)~\cite{paulraj2004overview} in microwave systems to massive MIMO~\cite{heath2016overview} in millimeter-wave (mmWave) systems and even ultra-massive MIMO (UM-MIMO)~\cite{akyildiz2016realizing} in terahertz (THz) systems, multi-antenna technologies, by leveraging the degrees of freedom (DoFs) in the spatial domain, have been pursuing higher data rates and reliability to meet the demand of the future-generation wireless communications.

\par In contrast to conventional multi-antenna systems that contain antennas at fixed positions, the concept of position-flexible antennas is proposed to further utilize the spatial variation of wireless channels in a given region without increasing the number of antennas~\cite{wong2020fluid,wong2021fluid,wong2022bruce,wong2022fluid,zhu2024movable,zheng2024flexible}. The flexibility in antenna position inside a spatial region improves the communication performance by adjusting the antenna to where is less impacted by the channel fading. Specifically, fluid antenna system (FAS)~\cite{wong2020fluid,wong2022bruce,wong2022fluid} and movable antennas (MAs)~\cite{zhu2024movable,ma2024mimo} are two implementations of position-flexible antennas. On one hand, the fluid antenna technology, also regarded as liquid antennas, applies software-controllable liquid materials and features the flexibility and reconfigurability in its shape and position~\cite{martinez2022towards}. On the other hand, MAs refer to antennas with the ability of physical motion in local, assigned regions~\cite{zhuravlev2015experimental,li2022using}.

\par Till date, analytical channel models for position-flexible antenna systems have been developed~\cite{wong2021fluid,zhu2024modeling}. In~\cite{zhu2024modeling}, a deterministic channel model is proposed based on the superposition of multi-path components. The large amount of model parameters supports various analytical evaluation of the MA system, whereas the high complexity makes it difficult for the implementation. By contrast, authors in~\cite{wong2021fluid} proposed a spatially-correlated Rayleigh fading channel model, based on parameters that represent the analytical correlation between channels. The assumption of Rayleigh fading channels results in the real-valued covariance, which however is invalid for the complex-valued covariance of channel coefficients characterized from the measurement.
Moreover, in order to improve the performance and unlock the potential of position-flexible antenna systems, algorithms have been studied for antenna position selection~\cite{ma2024mimo,zhu2024multiuser,hu2024fluid,pi2023multiuser,chai2022port,ye2024fluid,mei2024movable}, beamforming design~\cite{zhu2023movable,ma2024multi,kang2024deep}, and joint optimization~\cite{chen2023joint,qin2024antenna,hu2024movable}.
Nevertheless, experimental evaluation of movable antenna systems with antenna position selection and beamforming is still absent.

\par To fill this research gap, in this paper, we provide a complete and practical assessment of movable antenna systems in the real-world environment, including the channel measurement, the channel model parameterized from measurement results, the port selection algorithm, and the beamforming design.
Specifically, we first develop a wideband channel measurement platform with a 0.2~mm-precision two-dimensional movable antenna system. Extensive channel measurement is carried out across $32\times32$ ports at 300~GHz. The variation of practical wireless channels across the different ports is elaborated in detail, by analyzing the line-of-sight (LoS) and the reflected transmission, as well as the multi-path fading attributed to their superposition. The practical two-ray model for each port is constructed and verified with the analytical result.
In light of the measurement results, spatial-correlated channel models for the two-dimensional MA system are proposed, statistically parameterized by the complex covariance matrix of measured channels.

\par Channel models are fundamentals for system design and performance evaluation. To this end, by distributing the given ports into uniform regions for each MA and maximizing the signal-to-interference-and-noise ratio (SINR), the antenna position selection algorithm is proposed, which is applicable to both planar and linear MA systems.
Finally, the beamforming algorithm is conducted and the performance, in terms of spectral efficiency, of different MA systems is evaluated and compared.
By coping with the multi-path fading, MA systems with the proposed uniform-region SINR-optimized position selection algorithm can improve the spectral efficiency by 11.48\% across the $32\times32$~mm$^2$ area in the THz wireless channel. Besides, in view of the channel characteristic and the position selection algorithm, $N\times 1$-type MA, better than $1\times N$ or $N\times N$ types, can reach 99.57\% of the optimal spectral efficiency obtained by traversing all candidate ports.

\par The remainder of this paper is organized as follows.
The wideband channel measurement with the MA system is introduced in Section~\ref{section: campaign}. The measurement results and the practical two-ray model are also elaborated in this section. Then, in light of the measurement result, we propose the parameterized spatial-correlated channel model based on channel characterization of the complex covariance matrix in Section~\ref{section: model}. To evaluate and compare the performance of MA systems, the uniform-region SINR-optimized antenna selection algorithm and the beamforming algorithm is performed for the MA system in Section~\ref{section: results}. Finally, the paper is concluded in Section~\ref{section: conclusion}.
\section{Channel Measurement Campaign} \label{section: campaign}

In this section, we first introduce the vector network analyzer (VNA)-based channel measurement system with the movable antenna system at Tx. Then we describe the measurement conducted in a small anechoic chamber at 300~GHz. Finally, measurement results are discussed and the two-ray channel model is constructed for 1024 ports based on properties of the LoS ray and the reflected ray.

\begin{figure}
    \centering
    \subfigure[The diagram of the measuring system with the movable antenna system.]{
    \includegraphics[width=\linewidth]{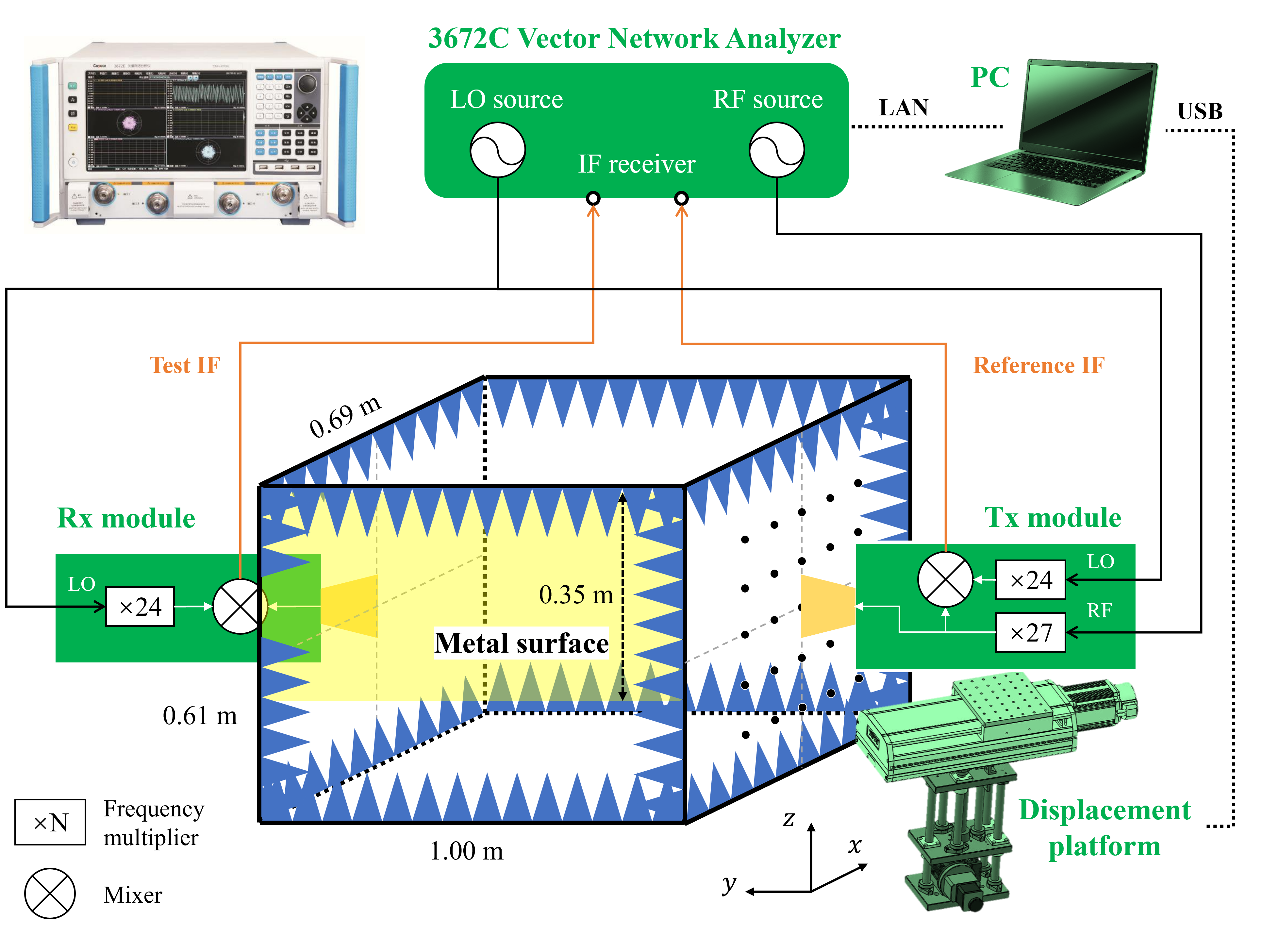}
    }
    \\
    \subfigure[The measurement scenario with the metal surface.]{
    \includegraphics[width=\linewidth]{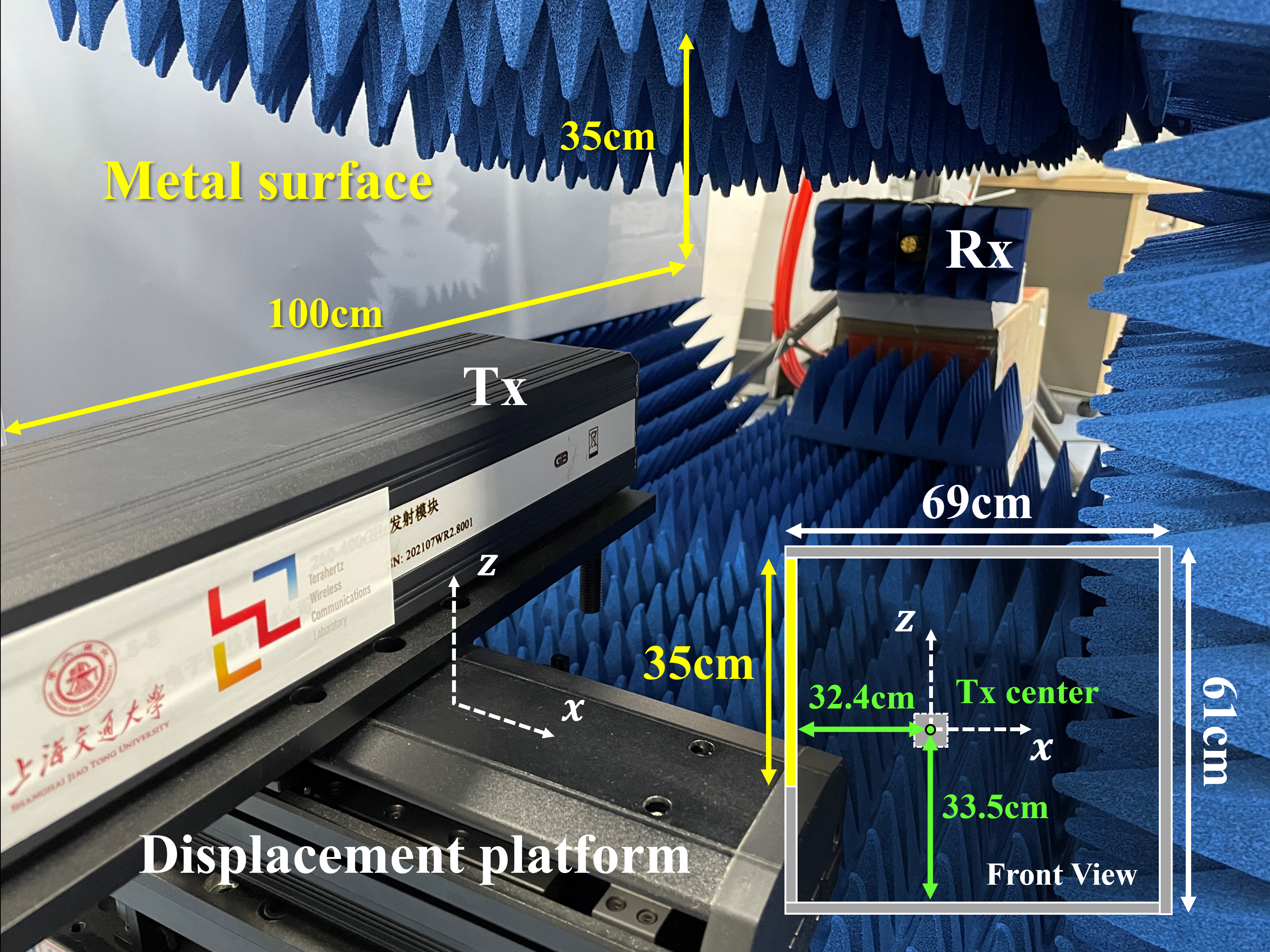}
    }
    \caption{Overview of the measurement platform for MA systems and the measurement scenario.}
    \label{fig:platform}
\end{figure}

\begin{table}
  \centering
  \caption{Parameters of the measurement system.}
    \setlength{\tabcolsep}{6mm}{
    \begin{tabular}{ll}
    \toprule
    Parameter & Value \\
    \midrule
    Frequency band              & 260-320~GHz \\
    Bandwidth                   & 60~GHz \\
    Time resolution             & 16.7~ps \\
    Space resolution            & 5~mm \\
    Sweeping interval           & 60~MHz \\
    Sweeping points             & 1001 \\
    Maximum excess delay        & 16.7~ns \\
    Waveguide gain              & 7~dBi\\
    Tx-Rx distance (aligned)    & 0.86~m \\
    Displacement precision      & 0.02~mm\\
    MA port type                & $32\times32$\\
    Port spacing                & 1~mm\\
    \bottomrule
    \end{tabular}
    }
  \label{tab:system_parameter}
\end{table}

\subsection{Measurement Deployment}
\subsubsection{Measurement platform}
\par The channel measurement platform is composed of three parts, as shown in Fig.~\ref{fig:platform}(a), including the computer (PC) as the control system, the displacement system, and the measuring system.

\par The measuring system consists of transmitter (Tx) and receiver (Rx) modules and the Ceyear 3672C VNA.
The VNA generates radio frequency (RF) and local oscillator (LO) sources. RF and LO signals are multiplied by 27 and 24, respectively in the THz module. 
The transmit and received RF signals are mixed by the LO signal and down-converted to the reference intermediate frequency (IF) signal at Tx and the test IF signal at Rx, respectively.
Two IF signals are sent back to the VNA, and the transfer function of the device under test (DUT), including the wireless channel and the system hardware is calculated as the ratio of the two frequency responses. To eliminate the influence of the system hardware, the calibration is conducted, which is described in detail in our previous works~\cite{wang2024far, wang2022thz, li2022channel}.

\par On one side, the VNA-based channel measurement system features extremely large bandwidth up to tens of GHz, which results in a high temporal resolution as large as tens of ps. On the other side, the measurement of MA systems requires the physical movement of antennas with a high spatial resolution. In the platform, the Tx is installed on a displacement system, which supports two-dimensional antenna movements in the $x$-$z$ plane perpendicular to the Tx-Rx line-of-sight at the precision of 0.02~mm. The $x$ and $z$ axes correspond to the horizontal and vertical movement, respectively.

\par The PC alternately controls the movement of Tx through the displacement system and the measuring process through the VNA.
The measurement starts from the element at the left bottom corner, and all ports are scanned first horizontally (in the $x$-axis) and then vertically (in the $z$-axis). The measuring time for each position is about 2.3~s.

\subsubsection{Measurement setup}
\par In this measurement, we investigate the THz frequency band ranging from 260~GHz to 320~GHz, which covers a substantially large bandwidth of 60~GHz. As a result, the resolution in time is 16.7~ps, corresponding to the resolution of propagation path length equal to 5~mm, which can distinguish MPCs that are close in arrival time.
The frequency sweeping interval is 60~MHz, resulting in 1001 sample points at each Tx-Rx position and equivalently, the maximum detectable path length of 5~m, which is sufficient for the small-scale measurement and saves the measuring time in return.
Key parameters of the measurement are summarized in Table~\ref{tab:system_parameter}.
\par As shown in Fig.~\ref{fig:platform}(b), the wireless propagation channel is confined in a 0.69~m long, 0.61~m wide, and 1.00~m deep anechoic chamber. A metal surface is vertically attached to one side surface of the chamber. Other inner surfaces are wrapped by absorbing materials to restrain the multi-path effect.
Tx and Rx are placed at two ends of the anechoic chamber.

\par In this measurement, $32\times32$ ports for the Tx antenna are measured with the spacing of 1~mm.
The Rx is static and aligned with the center of selectable ports at the Tx, whose distance is 0.86~m.
The measuring time is subject to the motion driven by the displacement platform, which takes about 1 hour to traverse 32$\times$32 ports.

\subsection{Measurement Results}
Measurement results are discussed next, including the gain and the phase of the LoS ray and the metal surface-reflected ray across the $32\times32$ ports.
Besides, the multi-path fading caused by the superposition of the two rays is elaborated.

\subsubsection{The LoS and the reflected ray}

\begin{figure}
    \centering
    \includegraphics[width=\linewidth]{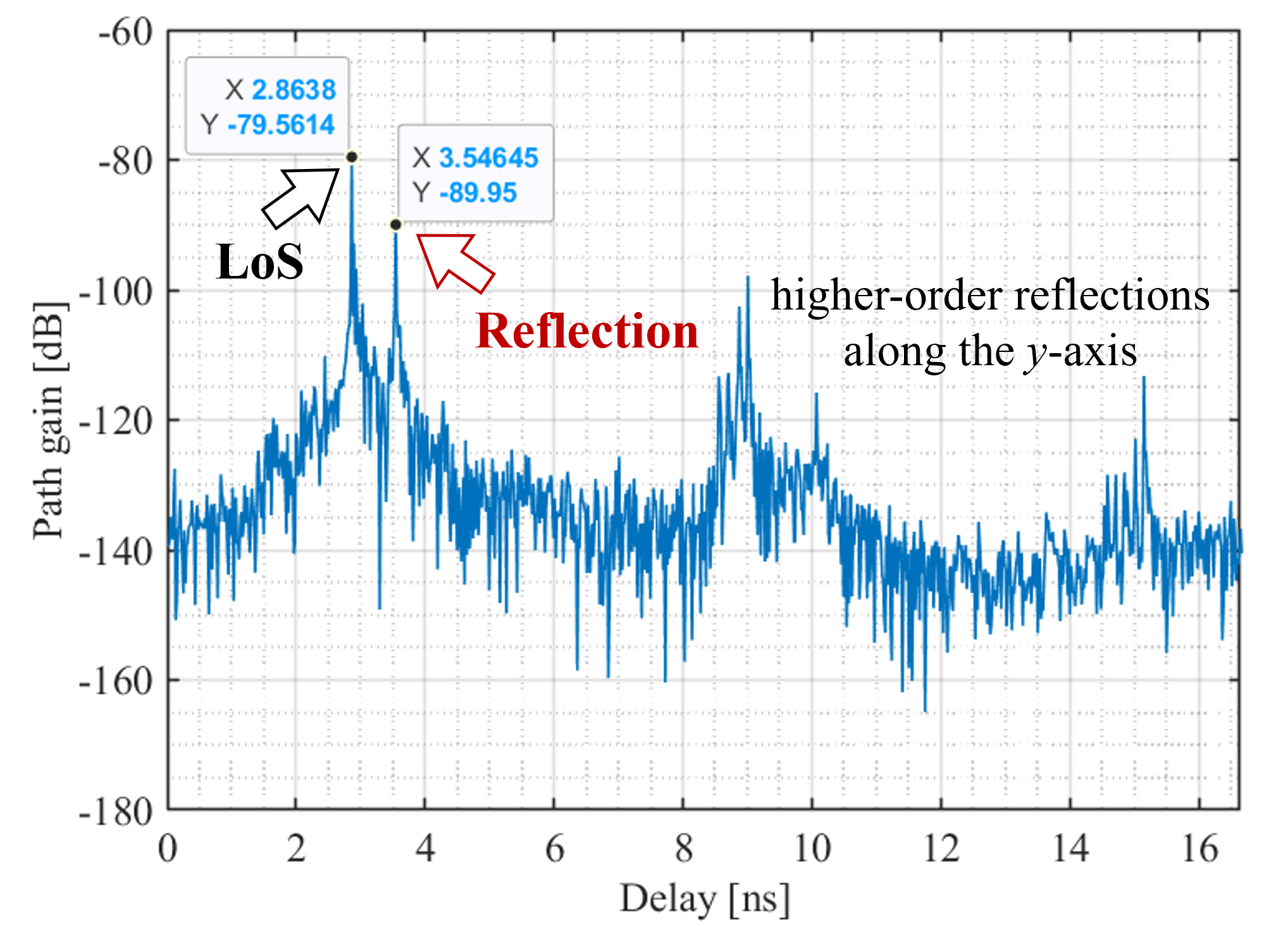}
    \caption{The CIR result at the center of Tx antenna ports.}
    \label{fig:CIR_UPA_center}
\end{figure}

\par In the measurement, for each channel from one port at Tx to the Rx, the frequency-domain measurement derives 1001 samples in channel transfer function (CTF), which is transferred into the channel impulse response (CIR) by inverse Fourier transform (IFT).
The CIR result from the center of the Tx antenna to the Rx is shown in Fig.~\ref{fig:CIR_UPA_center}. Due to the high resolution of 16.7~ps, we can distinguish multi-path components in the time domain from the CIR result. 

\par The LoS ray arrives at 2.8638~ns, corresponding to the traveling distance of 0.8591~m, and has the greatest path gain of -79.6~dB. 
Right after the LoS rays comes the major reflected ray from the metal surface at 3.54645~ns, which travels 1.0639~m and has a second-greatest path gain of -89.95~dB.
Besides, since the two ends of the chamber are not covered by absorbing materials, we can also observe higher-order reflected rays. These rays are separated by 6~ns, corresponding to a back-and-forth distance between the Tx and the Rx along the $y$-axis inside the chamber.

\begin{figure}
    \centering
    \subfigure[Path gain of the LoS ray.]{
    \includegraphics[width=0.98\linewidth]{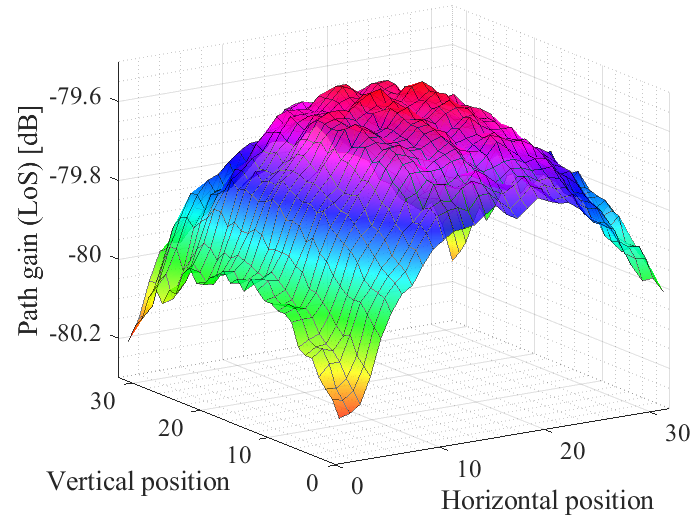}
    }
    \\
    \subfigure[Phase of the LoS ray.]{
    \includegraphics[width=0.98\linewidth]{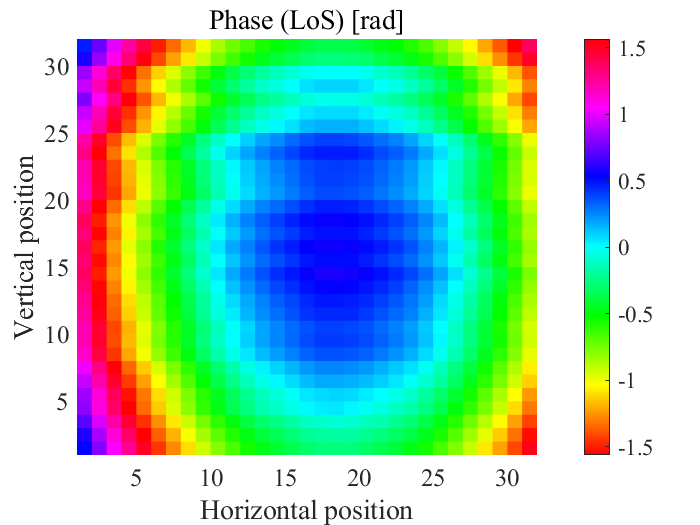}
    }
    \caption{Gain and phase change of the LoS ray across $32\times32$ port positions.}
    \label{fig:gain_phase_result_los}
\end{figure}

\begin{figure}
    \centering
    \subfigure[Path gain of the reflected ray.]{
    \includegraphics[width=0.98\linewidth]{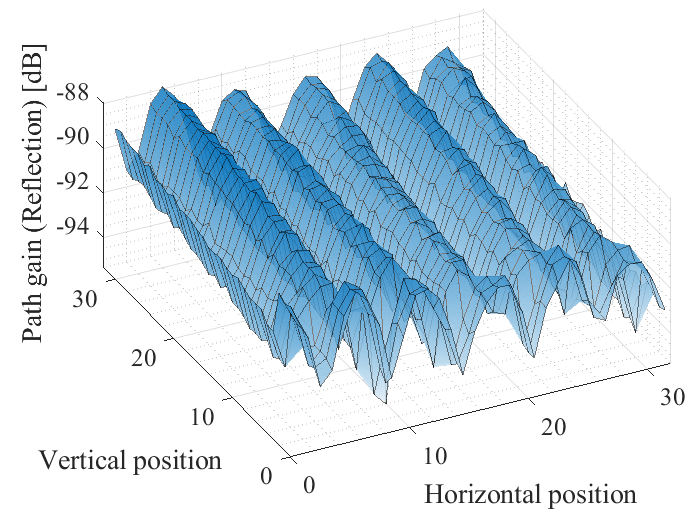}
    }
    \\
    \subfigure[Phase of the reflected ray.]{
    \includegraphics[width=0.98\linewidth]{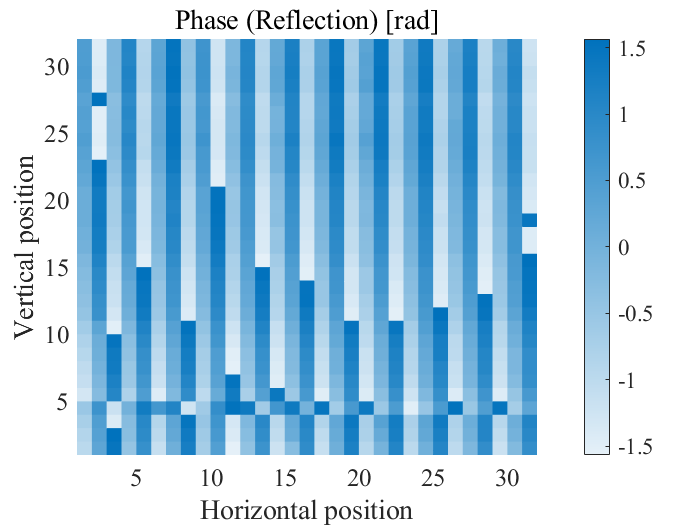}
    }
    \caption{Gain and phase change of the reflected ray across $32\times32$ port positions.}
    \label{fig:gain_phase_result_reflect}
\end{figure}

\par The change of properties of the LoS ray across all port positions is shown in Fig.~\ref{fig:gain_phase_result_los}.
Starting from the center, properties of the LoS ray change evenly along all directions as the position becomes away from the center. The path gain decreases by 0.6686~dB and the phase changes by $\pi$ from the center to the farthest corner.

\par The change of properties of the major reflection ray across all port positions is shown in Fig.~\ref{fig:gain_phase_result_reflect}.
The properties of the reflected ray vary differently along two axes. Specifically, the change along the $x$-axis is much more rapid than the counterparts along the $z$-axis. This is attributed to the scenario where the metal reflection surface is vertically placed on one side of the channel.
Specifically, the path gain changes periodically between -95~dB and -88~dB among horizontal positions along the $x$-axis. The period between two adjacent maxima is 6~mm.
By contrast, the path gain barely changes among vertical positions, as the path gain difference is less than 1~dB along the $z$-axis.
The phase of the reflected ray changes linearly along both axes, whereas the change rate is much larger between horizontal positions than between vertical positions.

\subsubsection{Multi-path fading and the two-ray model}

\begin{figure}
    \centering
    \includegraphics[width=\linewidth]{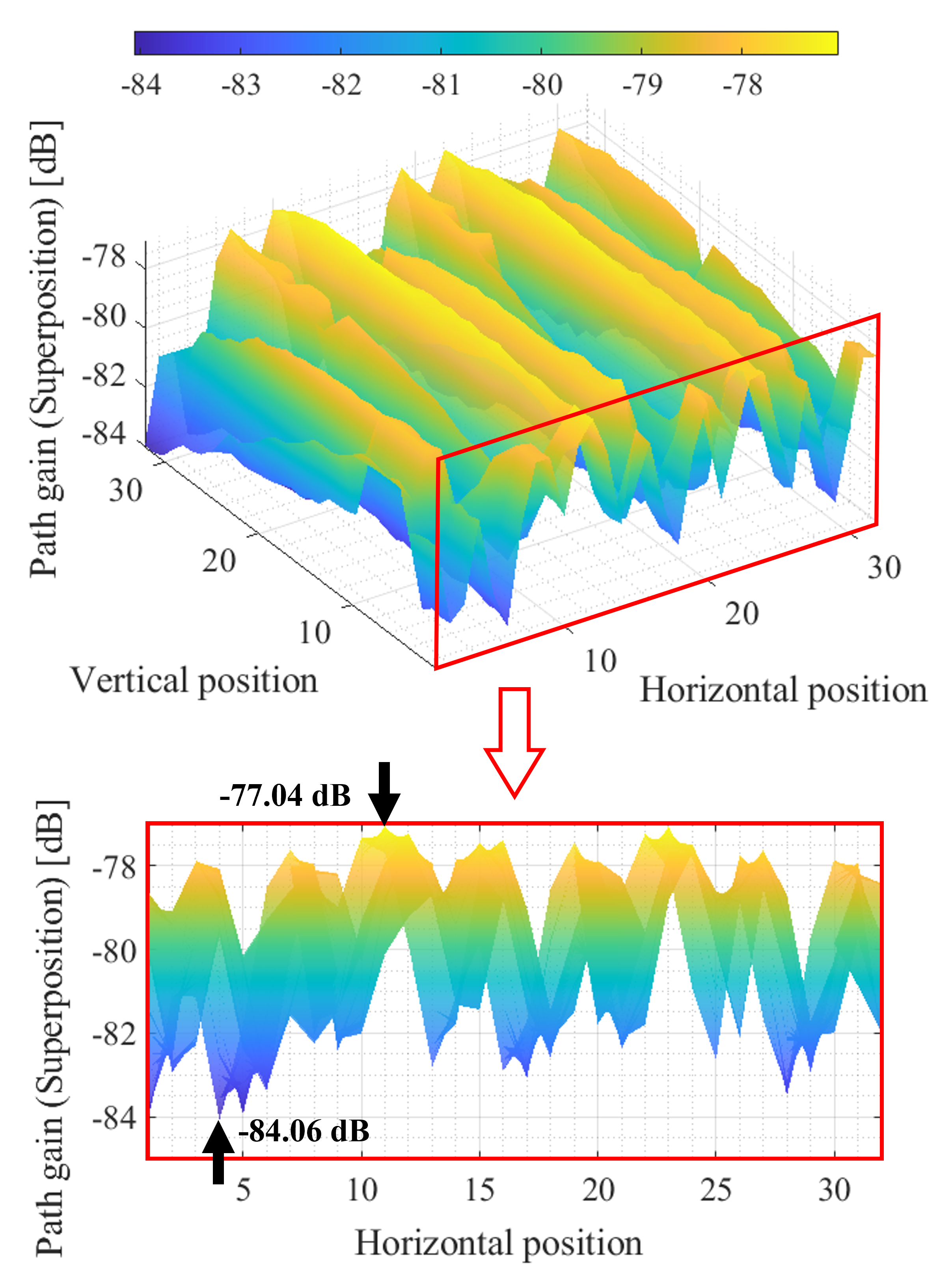}
    \caption{Superposition of the LoS ray and the reflected ray.}
    \label{fig:superposition_result}
\end{figure}

\begin{figure}
    \centering
    \includegraphics[width=\linewidth]{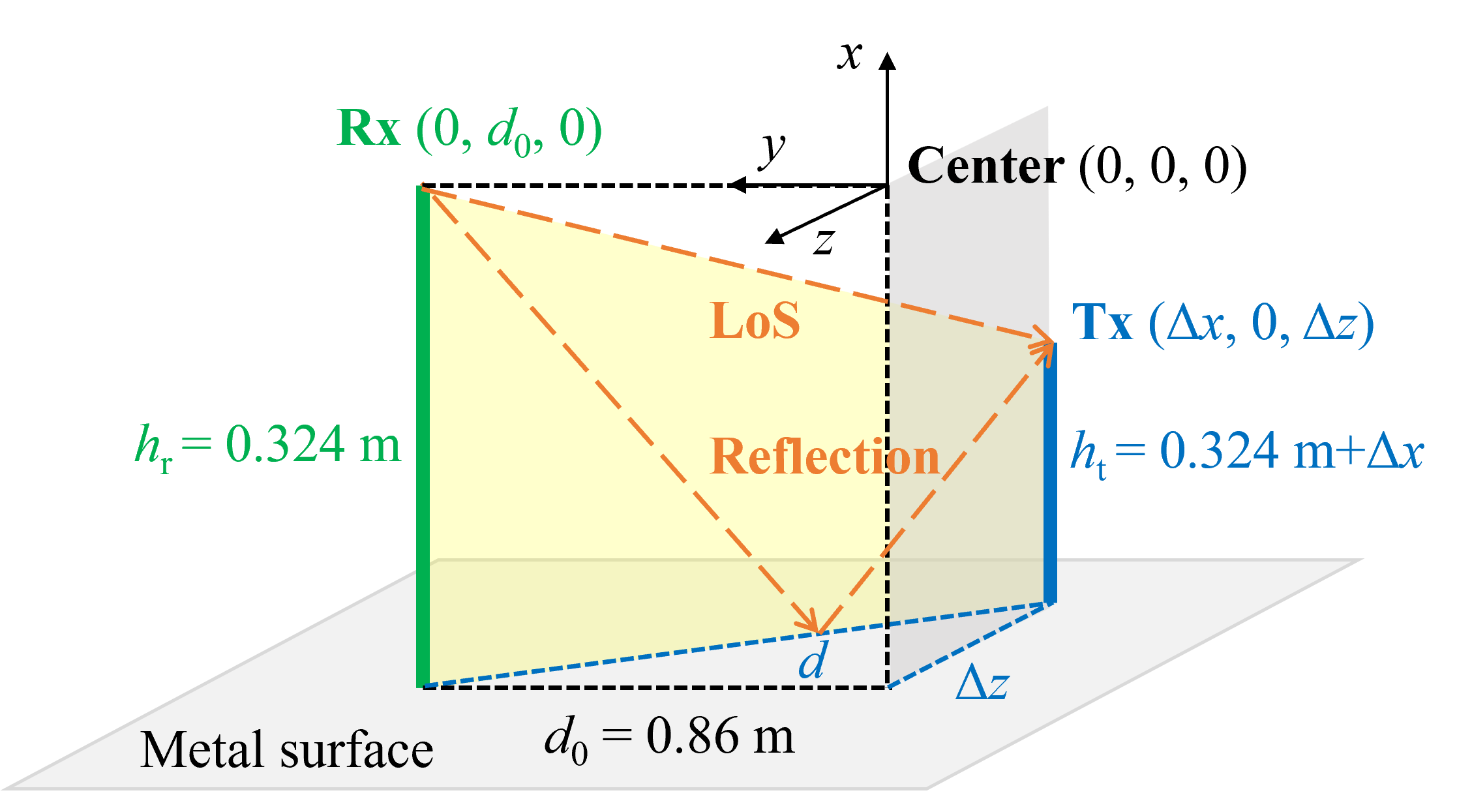}
    \caption{The two-ray model for the measurement scenario.}
    \label{fig:diagram_two_ray_model}
\end{figure}

\par The magnitude and phase of the LoS ray and the major reflected ray are extracted, and then we construct a two-ray channel model as follows,
\begin{equation}
\begin{aligned}
    h(\tau,f) &= h_{\rm LoS}(\tau,f)+h_{\rm reflect}(\tau,f)\\
    &=\alpha_{\rm LoS}(f)\delta(\tau-\tau_{\rm LoS}) + \alpha_{\rm reflect}(f)\delta(\tau-\tau_{\rm reflect}),
\end{aligned}
\end{equation}
where $\alpha_{\rm LoS}$ and $\alpha_{\rm reflect}$ denotes the complex gain of the LoS and the major reflected ray, respectively. $\tau_{\rm LoS}$ and $\tau_{\rm reflect}$ represents the arrival time.
\par The result is illustrated in Fig.~\ref{fig:superposition_result} and discussed as follows. Clearly, a sequence of power maxima and minima is observed along the $x$-axis among horizontal positions.
The classic two-ray model considers the wireless channel that contains the LoS ray and the ground-reflected ray. As shown in Fig.~\ref{fig:diagram_two_ray_model}, by regarding the metal surface as the ``ground'', the constructed model derived in the chamber can be translated into the classic two-ray model. As a result, the ``horizontal'' separation of the Tx-Rx antennas is $d=\sqrt{{d_0}^2+{\Delta z}^2}$, the receiver ``height'' is $h_{\rm r}=0.324$~m, and the transmitter ``height'' is $h_{\rm t}=\Delta x+0.324$~m. 
According to the simulation result of the classic two-ray model, multi-path fading occurs as $h_{\rm t}<d<4h_{\rm t}h_{\rm r}/\lambda$, which accords with the measurement result. To be concrete, in this measurement scenario, $d$ is dominant by $d_0=0.86$~m, as the horizontal and vertical movements $\Delta x$ and $\Delta z$ are mm-level. Besides, $d<4h_{t}h_{r}/\lambda$ is valid since $\lambda$ is as small as 1~mm at 300~GHz. Therefore, a sequence of power maxima and minima can be observed in the constructed two-ray model derived from the measurement result.


\section{Spatial-Correlated Channel Models for Two-Dimensional Movable Antenna Systems} \label{section: model}

In this section, we propose the spatial-correlated channel model for the two-dimensional MA system. Specifically, the model is parameterized by the covariance matrix of measured channels. We start by the characterization of the complex-valued covariance matrix of the channel coefficient and the real-valued covariance matrix of the path gain. The simulation results of path gain and channel coefficient obtained from the channel models are compared with the measurement results, respectively.
Note that the real and the imaginary parts of the channel coefficient are generated independently in the Rayleigh fading model~\cite{wong2021fluid}, which results in the real-valued covariance. However, we characterize the complex-valued covariance of channel coefficient from the measurement, and therefore, the proposed channel model directly generates complex, spatial-correlated channel coefficients.

\subsection{Characterization of Spatial Covariance Matrix}
\par On account of port positions that are closely packed within several wavelengths in the MA system, channel coefficients at these ports are correlated.
As depicted in Fig.~\ref{fig:superposition_result}, the multi-path fading happens between ports along horizontal positions, while the change of channels in vertical positions is comparably insignificant. Therefore, we denote complex channel coefficients at horizontal positions in $n$ rows as $\mathbf{H}=[H_1, H_2, ..., H_n]^\text{T}$. Channel coefficients at different vertical positions in the $i$-th horizontal position, as column elements in the $i$-th row, are regarded as samples of $H_i$.

\par We characterize their correlation by the complex covariance matrix $\mathbf{\Sigma}_{\mathbf{H}}$ as
\begin{equation}
    \mathbf{\Sigma}_{\mathbf{H}} =
    \begin{bmatrix}
        \text{Var}\{H_1\} & \text{Cov}\{H_1,H_2\} & \cdots & \text{Cov}\{H_1,H_n\} \\
        \text{Cov}\{H_2,H_1\} & \text{Var}\{H_2\} & \cdots & \text{Cov}\{H_2,H_n\} \\
        \vdots & \vdots & \ddots & \vdots \\
        \text{Cov}\{H_n,H_1\} & \text{Cov}\{H_n,H_2\} & \cdots & \text{Var}\{H_n\} 
    \end{bmatrix},
\end{equation}
where
\begin{equation} \label{eq:complex_cov}
\begin{aligned}
    \text{Cov}\{H_i, H_j\} &= \text{E}\{(H_i-E\{H_i\})(H_j-E\{H_j\})^*\} \\
    &= \text{E}\{H_i{H_j}^*\} - E\{H_i\} {E\{H_j\}}^*,
\end{aligned}
\end{equation}
is the covariance between complex channel coefficients $H_i$ and $H_j$. For $i=j$, the variance is expressed as
\begin{equation} \label{eq:complex_var}
\begin{aligned}
    \text{Var}\{H_i\} &= \text{E}\{(H_i-E\{H_i\})(H_i-E\{H_i\})^*\} \\
    &= \text{E}\{|H_i-E\{H_i\}|^2\}.
\end{aligned}
\end{equation}
Since $\mathbf{H}$ is complex, the operations are generalized for complex values, and the covariance matrix $\mathbf{\Sigma}_{\mathbf{H}}$ is complex-valued.

\subsection{Spatial-Correlated Model for Path Gain}

\par To start with, we can alternatively characterize the path gain, i.e. the magnitude of channel response $\mathbf{|H|}=[|H_1|, |H_2|, ..., |H_n|]^\text{T}$ by its real-valued and symmetric covariance matrix
\begin{equation}
    \mathbf{\Sigma}_{\mathbf{|H|}} =
    \begin{bmatrix}
        \text{Var}\{|H_1|\} & \cdots & \text{Cov}\{|H_1|,|H_n|\} \\
        \vdots & \ddots & \vdots \\
        \text{Cov}\{|H_n|,|H_1|\} & \cdots & \text{Var}\{|H_n|\} 
    \end{bmatrix}.
\end{equation}

\par The correlated real-valued random variables $\mathbf{|H|}$ can be modeled by
\begin{equation}\label{eq:channel_generate_real}
    \mathbf{|H|}=\mathbf{\mu}_{\mathbf{|H|}}+\mathbf{C}\mathbf{X},
\end{equation}
where $\mathbf{\mu}_{\mathbf{|H|}} = [\text{E}\{|H_1|\},\text{E}\{|H_2|\},...,\text{E}\{|H_n|\}]^\text{T}$ is the mean vector.

\par For a symmetric positive definite covariance matrix $\mathbf{\Sigma}$, the matrix $\mathbf{C}$ is given by the Cholesky decomposition such that
\begin{equation}
    \mathbf{\Sigma}_{\mathbf{|H|}}=\mathbf{C}\mathbf{C}^\text{T}.
\end{equation}
For a non-positive definite matrix which is symmetric, the decomposition is given by  $\mathbf{\Sigma}_{\mathbf{|H|}}=\mathbf{L}\mathbf{D}\mathbf{L}^\text{T}$ where $\mathbf{L}$ is a lower triangular matrix whose diagonal elements are 1 and $\mathbf{D}$ is a positive diagonal matrix. In this case, the matrix $\mathbf{C}$ is given by
\begin{equation}\label{eq:ldl}
    \mathbf{C} = \mathbf{L}\sqrt{\mathbf{D}}.
\end{equation}

\par The set of random variables $\mathbf{X} = [X_1,X_2,...,X_n]^\text{T}$ contains uncorrelated random variables $X_i$ ($i=1,2,...,n)$ that have zero mean and unit variance. The distribution function of $X_i$ determines the samples of $|H_i|$. In our measurement, the magnitude of channel response, $|g_{i,k}|$, is uniformly distributed in each row (for each $i=1,2,...,32$). Therefore, we apply
\begin{equation}\label{eq:Xi}
    X_i\sim\mathbb{U}[-\sqrt{3},\sqrt{3}].
\end{equation}

\begin{figure*}
    \centering
    \subfigure[Measurement result, $|g_{i,k}|.$]{
    \includegraphics[width=0.35\linewidth]{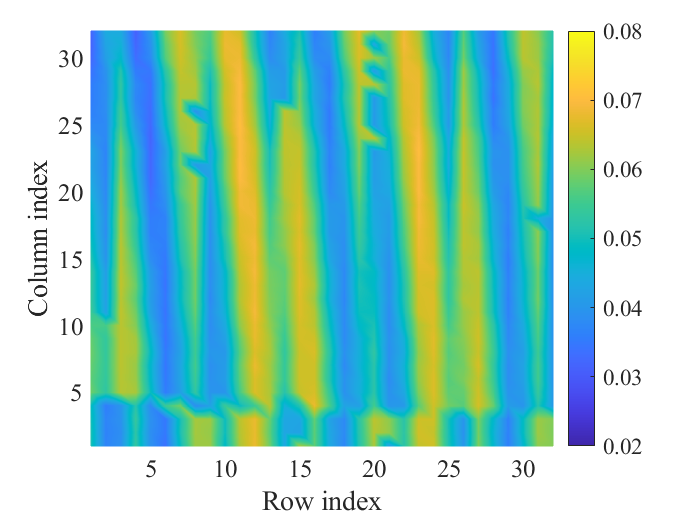}
    }
    \subfigure[CDF of $|g_{i,k}|$ for $i=1,2,...,32$.]{
    \includegraphics[width=0.25\linewidth]{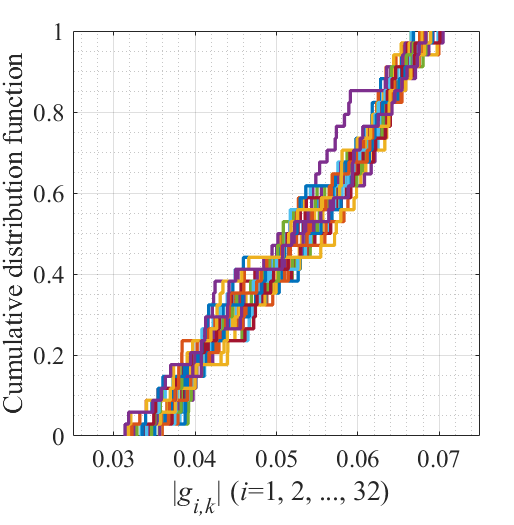}
    }
    \subfigure[Measurement result, $\mathbf{\Sigma_{|G|}}$.]{
    \includegraphics[width=0.33\linewidth]{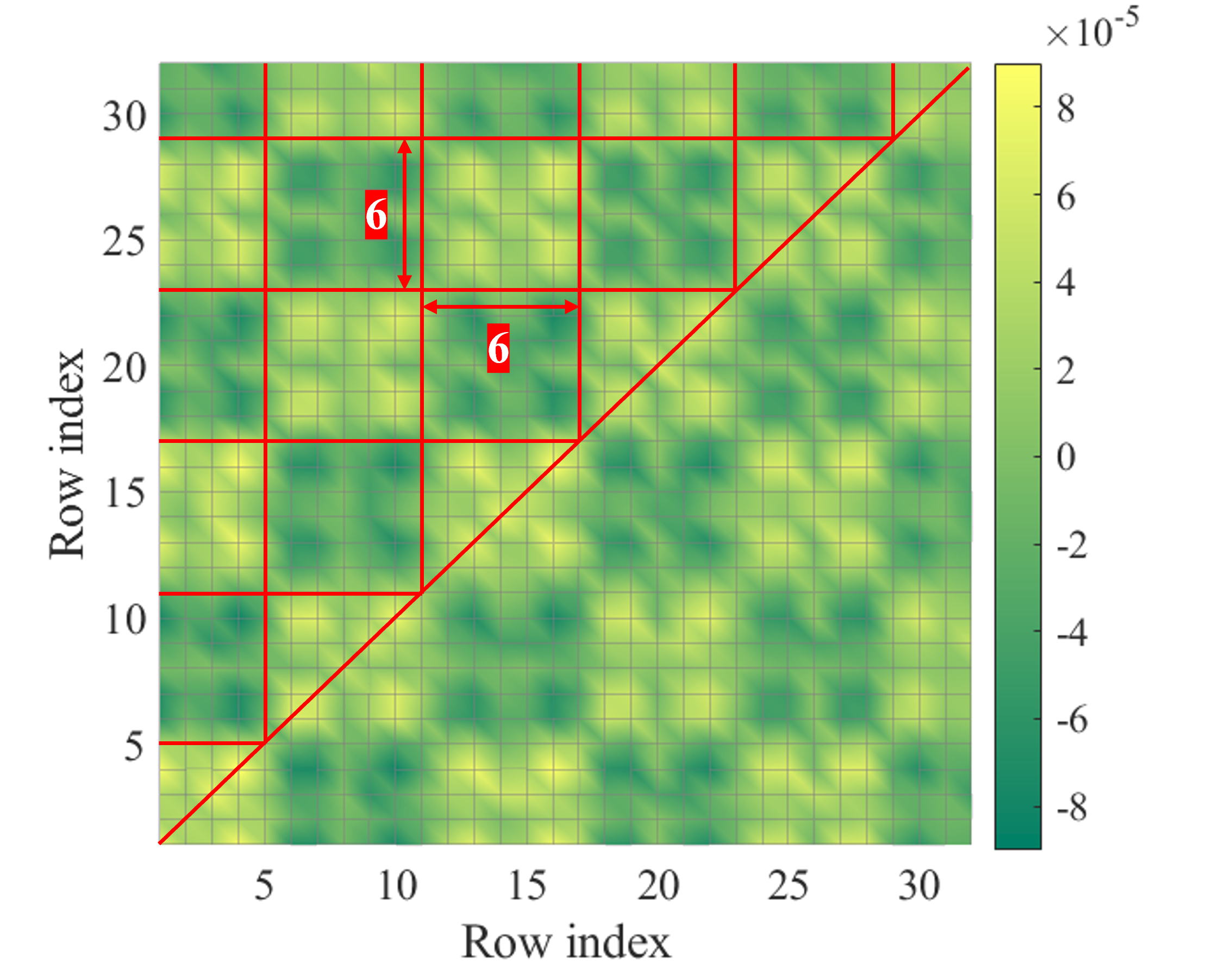}
    }
    \\
    \subfigure[Simulation result, $|h_{i,k}|$.]{
    \includegraphics[width=0.35\linewidth]{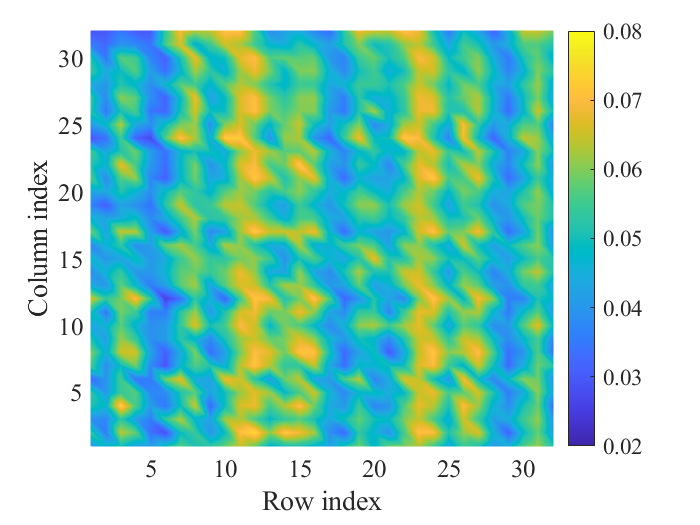}
    }
    \subfigure[CDF of $|h_{i,k}|$ for $i=1,2,...,32$.]{
    \includegraphics[width=0.25\linewidth]{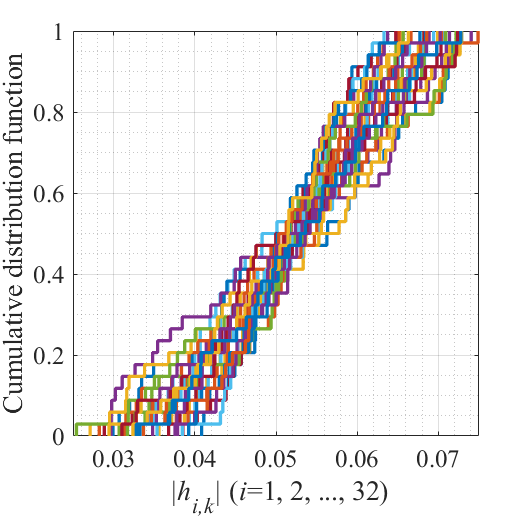}
    }
    \subfigure[Simulation result, $\mathbf{\Sigma_{|H|}}$.]{
    \includegraphics[width=0.33\linewidth]{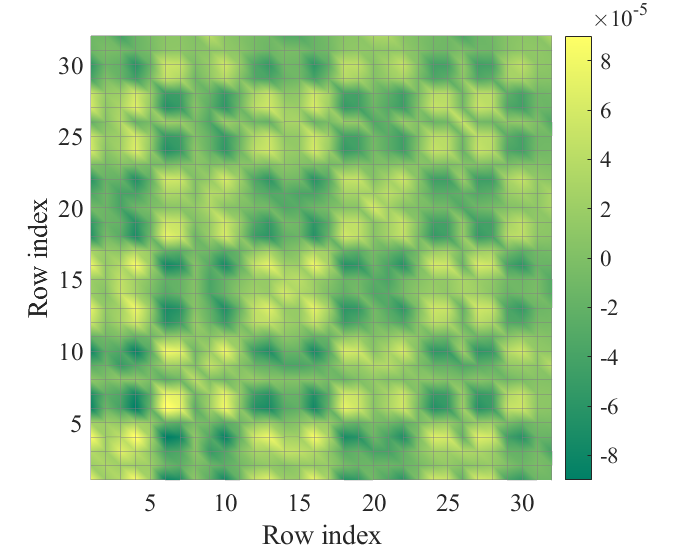}
    }
    \caption{Measurement and simulation result of channel response magnitude.}
    \label{fig:magnitude_sim_result}
\end{figure*}

\par
\begin{figure}
    \centering
    \includegraphics[width=\linewidth]{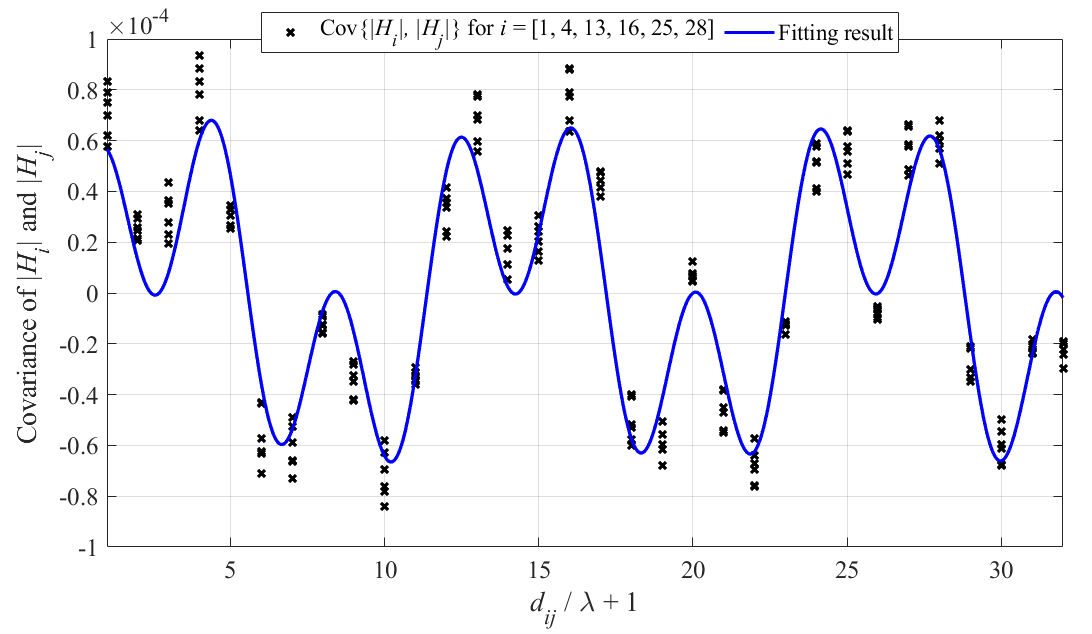}
    \caption{The periodic change of $\text{Cov}\{|H_{1}|,|H_{j}|\}$ with $d_{ij}$.}
    \label{fig:magnitude_cov_model}
\end{figure}

 The modeling result of the channel response magnitude at $32\times32$ ports is shown and compared with the measurement result in Fig.~\ref{fig:magnitude_sim_result}.
First, the uniform distribution $X_i$ \eqref{eq:Xi} only captures the distribution of $|g_{i,k}|$ for each $i=1,2,...,32$, but cannot reveal the spatial correlation at positions in adjacent columns. Therefore, in Fig.~\ref{fig:magnitude_sim_result}(d), the result given by the model is not continuous between columns in each row. Second, the covariance matrix of the measured channel gain $\mathbf{\Sigma_{|G|}}$ is well reproduced by the channel model \eqref{eq:channel_generate_real} in Fig.~\ref{fig:magnitude_sim_result}(f). For the two-ray channel, the covariance of $|H_{i}|$ and $|H_{j}|$ can be modeled by
\begin{equation}
\begin{cases}
    \begin{aligned}
    \text{Cov}\{|H_{i}|,|H_{j}|\} =& \text{Cov}\{|H_{p}|,|H_{q}|\} \\
    &\text{if } [d_{ip},d_{jq}] \in \{[6\lambda,6\lambda],[0,12\lambda]\}
    \end{aligned}
    \\
    \begin{aligned}
    \text{Cov}\{|H_{1}|,|H_{j}|\} =& a_1 \sin(b_1*(d_{1j}/\lambda+1)+c_1) \\
    +& a_2 \sin(b_2*(d_{1j}/\lambda+1)+c_2)
    \end{aligned}
\end{cases}
\end{equation}
where $d_{ij}$ denotes the distance between the $i$-th and the $j$-th row. Specifically, the covariance of magnitude between two rows varies periodically. The value is identical if one position is fixed and the other changes by $12\lambda$, or if two positions both change by $6\lambda$. Moreover, inside the first period, $\text{Cov}\{|H_{1}|,|H_{j}|\}$, can be modeled by the superposition of two sinusoidal functions as shown in Fig.~\ref{fig:magnitude_cov_model}. The parameters are obtained by fitting results, as $a_1 = 4.116\times10^{-5}$, $b_1 = 0.5468$, $c_1 = 0.004135$, $a_2 = 4.149\times10^{-5}$, $b_2 = 1.6160$, and $c_2 = 0.5212$.

\subsection{Spatial-Correlated Model for Channel Coefficients}

\begin{figure*}
    \centering
    \subfigure[Measurement result, Real\{$g_{i,k}$\}.]{
    \includegraphics[width=0.4\linewidth]{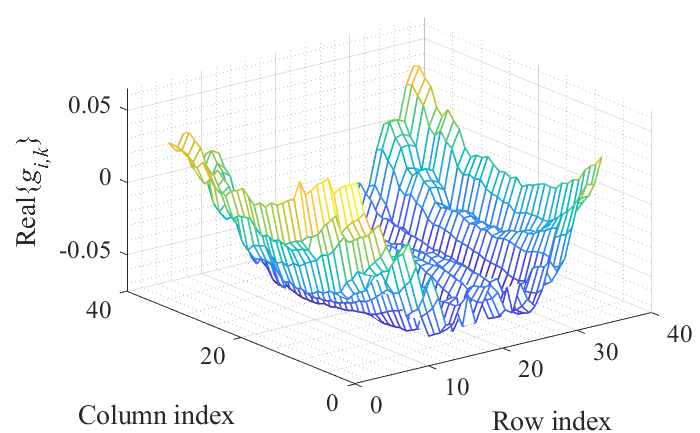}
    }
    \subfigure[Measurement result, Imag\{$g_{i,k}$\}.]{
    \includegraphics[width=0.4\linewidth]{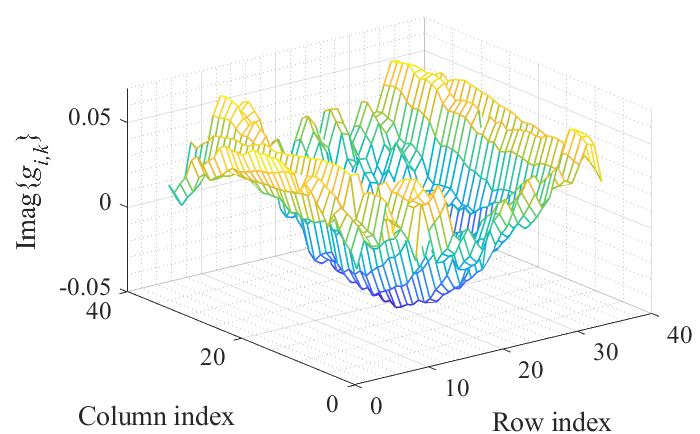}
    }
    \\
    \subfigure[Measurement result, Real\{$\mathbf{\Sigma_{G}}$\}.]{
    \includegraphics[width=0.4\linewidth]{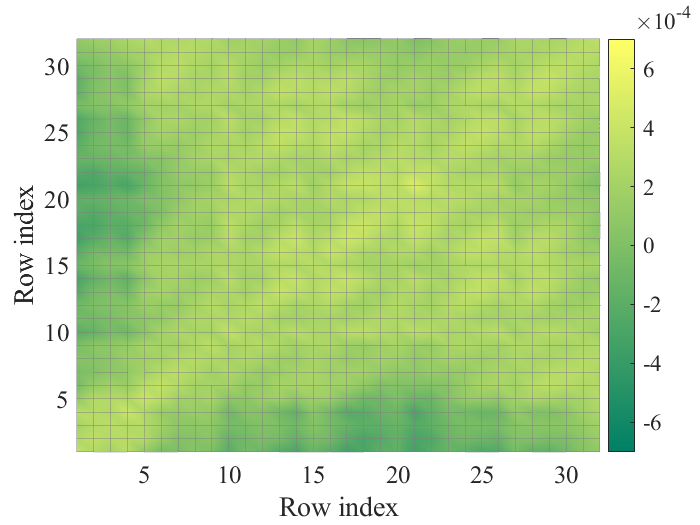}
    }
    \subfigure[Measurement result, Imag\{$\mathbf{\Sigma_{G}}$\}.]{
    \includegraphics[width=0.4\linewidth]{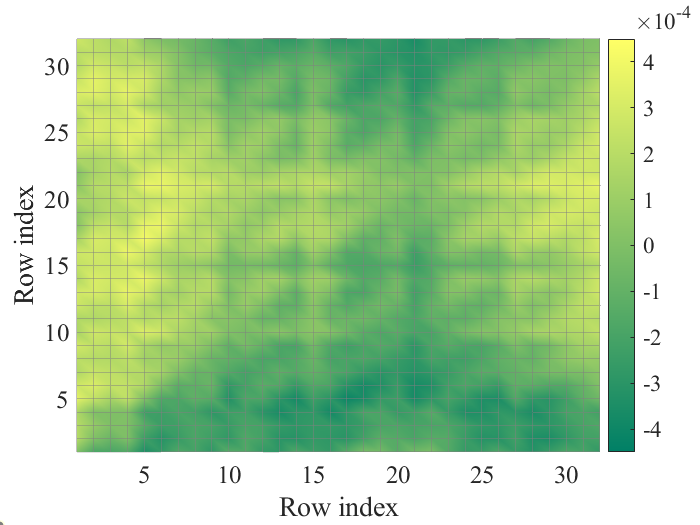}
    }
    \\
    \subfigure[Simulation result, Real\{$h_{i,k}$\}.]{
    \includegraphics[width=0.4\linewidth]{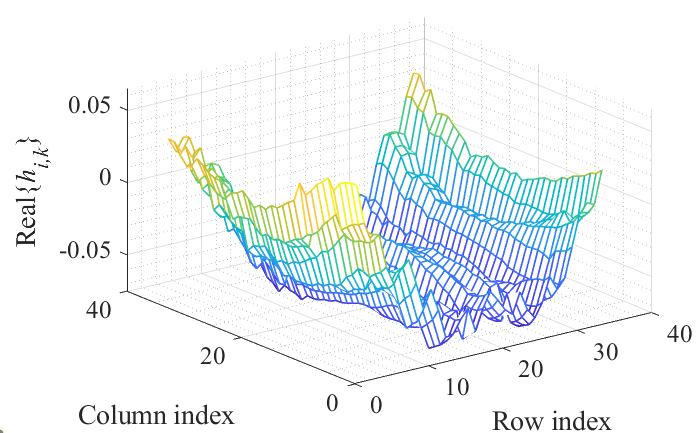}
    }
    \subfigure[Simulation result, Imag\{$h_{i,k}$\}.]{
    \includegraphics[width=0.4\linewidth]{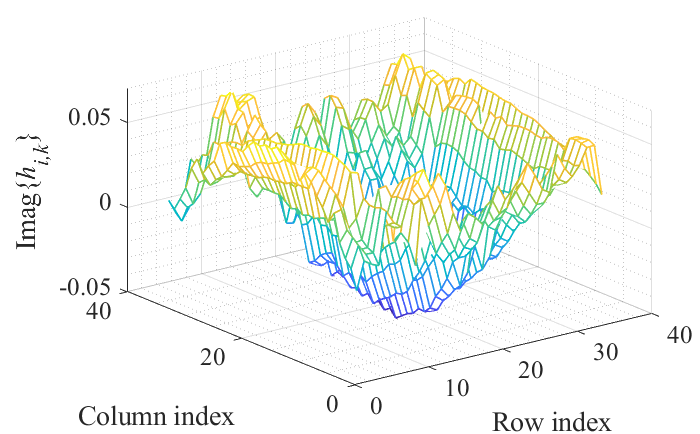}
    }
    \\
    \subfigure[Measurement result, Real\{$\mathbf{\Sigma_{H}}$\}.]{
    \includegraphics[width=0.4\linewidth]{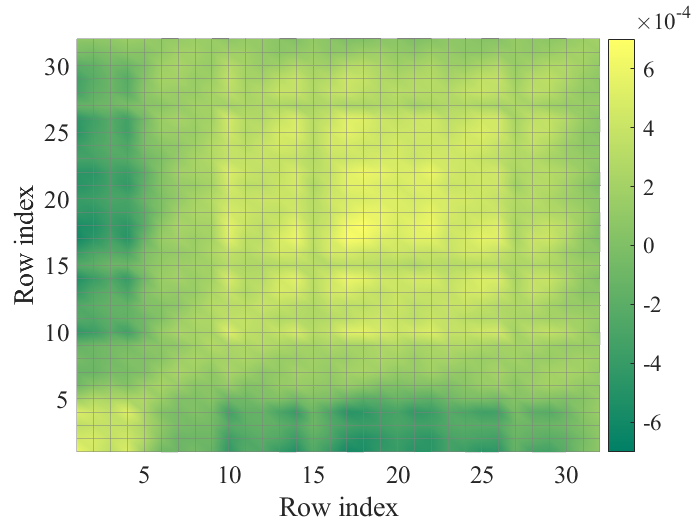}
    }
    \subfigure[Measurement result, Imag\{$\mathbf{\Sigma_{H}}$\}.]{
    \includegraphics[width=0.4\linewidth]{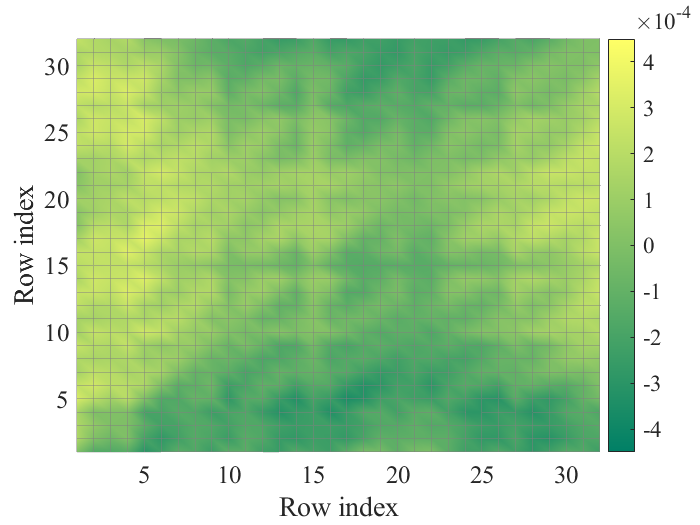}
    }
    \caption{Measurement and simulation result of complex channel coefficient.}
    \label{fig:coefficient_sim_result}
\end{figure*}

\par The method in \eqref{eq:channel_generate_real} and \eqref{eq:ldl} is also valid for the modeling of complex-valued channel coefficients, since the covariance matrix $\mathbf{\Sigma}_{\mathbf{H}}$ is complex-valued and symmetric. Specifically for two random variables, i.e., $H_i$ and $H_j$ ($i,j\in[1,n]$, $i\neq j$), \eqref{eq:ldl} is solved by
\begin{subequations}
\begin{align}
    \mathbf{L}
        &= 
        \begin{bmatrix}
        1 & 0 \\
        \frac{\text{Cov}\{H_i,H_j\}}{\text{Var}\{H_i\}} & 1
        \end{bmatrix} 
        = 
        \begin{bmatrix}
        1 & 0 \\
        \rho_{i,j}\sigma_j/\sigma_i & 1
        \end{bmatrix}, \\
    \mathbf{D}
        &=
        \begin{bmatrix}
        \text{Var}\{H_i\} & 0 \\
        0 & \text{Var}\{H_j\}-\frac{\text{Cov}\{H_i,H_j\}}{\text{Var}\{H_j\}}
        \end{bmatrix} \\
        &=
        \begin{bmatrix}
        \sigma_i^2 & 0 \\
        0 & \sigma_j^2(1-\rho_{i,j}^2)
        \end{bmatrix}, \\
    \mathbf{C} &= 
        \begin{bmatrix}
        \sigma_i & 0 \\
        \sigma_j\rho_{i,j} & \sigma_j\sqrt{1-\rho_{i,j}^2}
        \end{bmatrix},
\end{align}
\end{subequations}
denoting the complex variance by $\sigma_i^2 = \text{Var}\{H_i\}$ as expressed in \eqref{eq:complex_var} and $\rho_{i,j}=\text{Cov}\{H_i,H_j\}/(\sigma_i\sigma_j)$ where the complex covariance is calculated in \eqref{eq:complex_cov}.
Therefore, the correlated channel coefficient can be generated by
\begin{subequations} \label{eq:channel_generate_complex_1D}
\begin{align}
    H_i &= \mu_i + \sigma_i Y_i, \ i\in[1,n], \\
    H_j &= \mu_j + \sigma_j\rho_{i,j} Y_i + \sigma_j\sqrt{1-\rho_{i,j}^2} Y_j, \ j\neq i,
\end{align}
\end{subequations}
where $\mu_i=\text{E}\{H_i\}$ is the complex mean of $H_i$. Uncorrelated random variables $Y_1,Y_2,...,Y_n$ generates samples with zero mean and unit variance.

\par Note that if the real and the imaginary parts of the channel coefficient are independently modeled~\cite{wong2021fluid}, the covariance of complex channel coefficients would degrade into real values, and satisfies
\begin{equation}
\begin{aligned}
    \text{Cov}\{H_i,H_j\} =& \text{ Cov}\{\text{Real}(H_i),\text{ Real}(H_j)\} \\
    +& \text{ Cov}\{\text{Imag}(H_i),\text{ Imag}(H_j)\}.
\end{aligned}
\end{equation}
However, for channel characterization of the spatial covariance from measured samples, the covariance matrix, as one input parameter of the channel model, remains complex in most cases. Therefore, we take the complex channel coefficient as one individual variable, instead of modeling its real and the imaginary parts separately.

\par To reveal the simulated channel coefficients in both rows and columns, we generate 2D channel coefficients $h_{i,k}$ by regarding the $k$-th sample of $H_i$ as the channel coefficient at the $i$-row and the $k$-th column, i.e.,
\begin{subequations}
    \begin{align}
    h_{i,k} =& \mu_i + \sigma_i y_{i,k}, \ i\in[1,n],\ k\in[1,m], \\
    h_{j,k} =& \mu_j + \sigma_j\rho_{i,j} y_{i,k} + \sigma_j\sqrt{1-\rho_{i,j}^2} y_{j,k}, \ j\neq i,\ k\in[1,m],
\end{align}
\end{subequations}
where
\begin{equation}
    y_{i,k} = \frac{ g_{i,k} - \frac{1}{m}\sum_{k=1}^{m}g_{i,k} }{ \sqrt{\frac{1}{m}\sum_{k=1}^{m}|g_{i,k}-\frac{1}{m}\sum_{k=1}^{m}g_{i,k}|^2} },
\end{equation}
are derived from measured channel coefficients in the $i$-th row and normalized to zero mean and unit variance. In the generalized channel model, $Y_i$ reveals the distribution of channel coefficients in the $i$-th row, and its samples can be regarded as channel coefficients in different columns at this row. As long as the distribution has zero mean and unit variance, the covariance of channel coefficients between rows holds.

\par The modeling result of the complex channel coefficient at $32\times32$ ports is shown and compared with the measurement result in Fig.~\ref{fig:coefficient_sim_result}. The simulation result reproduces the complex channel coefficients across the 2D ports, and the characteristics in terms of covariance matrix is reserved.
\section{Performance Analysis of the MA systems} \label{section: results}
\par In light of the multi-path fading observed in channels across $32\times32$ ports for Tx antennas to the fixed Rx, we employ the concept of a movable antenna system, which selects the physical positions of antenna elements with the best channel condition and thus improves the system performance without introducing a large number of antennas~\cite{zhu2024movable}. In this section, we first introduce the setup of antenna arrays at Tx and how we distribute the given selectable ports into movable regions and select positions for antenna elements. Then, we employ a beamforming algorithm and evaluate the performance, in terms of spectral efficiency, of MA systems.

\subsection{Movable Antenna Position Selection} \label{sec:port_select}

\subsubsection{Uniform-Region SINR-Optimized Position Selection}
\begin{figure*}
    \centering
    \includegraphics[width=\linewidth]{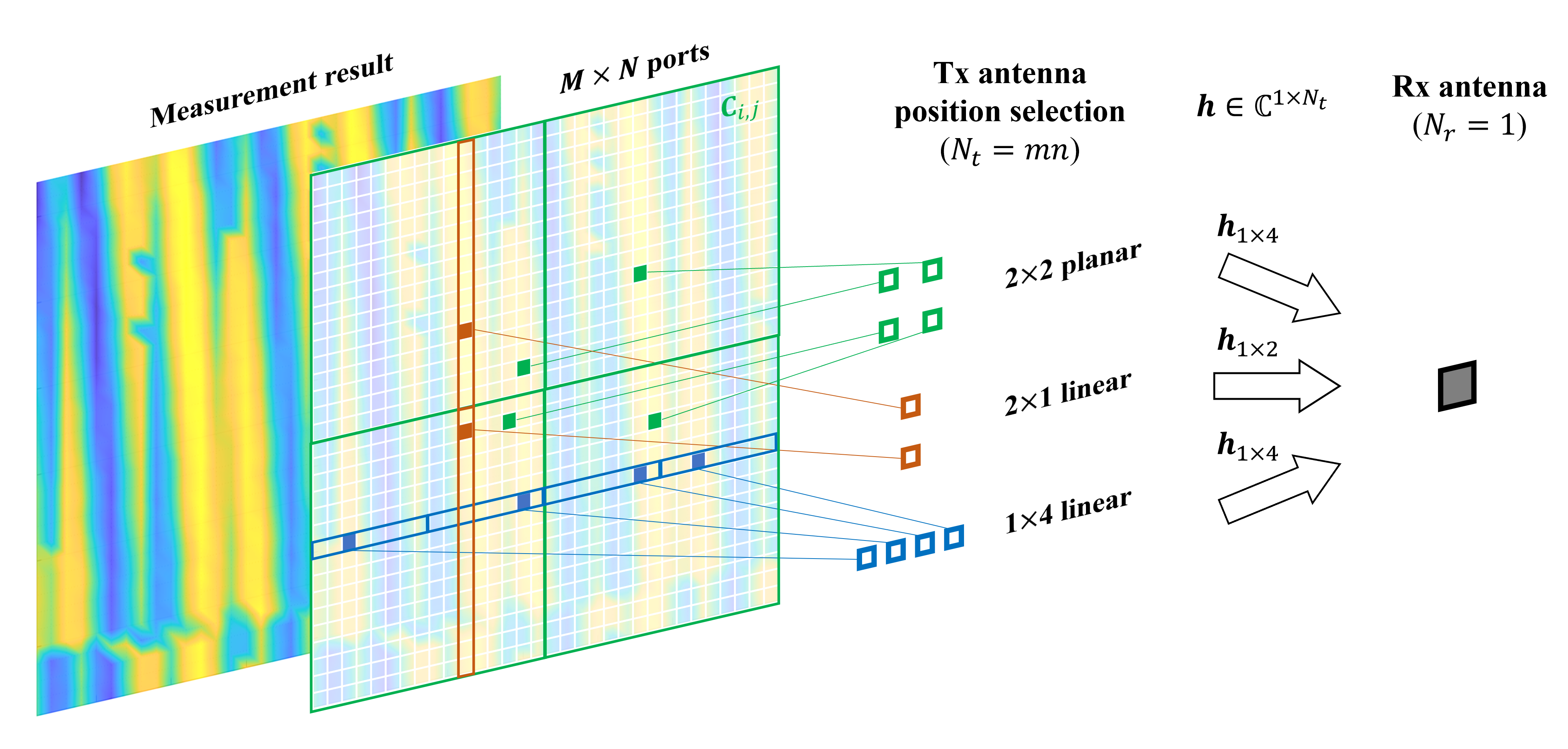}
    \caption{Antenna position selection for two-dimensional $m\times n$ MA systems at Tx out of $M\times N$ ports in the given area ($M=N=32$ in the measurement).}
    \label{fig:ma_select_port}
\end{figure*}

\par We simulate the Tx equipped with $m\times n$ planar MAs and $m\times1$ or $1\times n$ linear MAs. We elaborate the antenna position selection scheme in the given area $\mathbf{C}$ with $M\times N$ ports as follows. The set of candidate ports are represented by $\mathbf{C}=\{(x,y)\ |\ x\in[1,N], y\in[1,M]\}$. The coordinates of the transmit MAs are denoted by $\mathbf{t}=\left(\mathbf{t}_1, \mathbf{t}_2, ..., \mathbf{t}_{N_t}\right)^T$, where $\mathbf{t}_u = (x_{u}, y_{u}) \in \mathbf{C}$ for each $u=1,2,...,N_t$.

\par For the Tx equipped with $m\times n$ MAs, $M\times N$ ports from the measurement are uniformly divided into $m\times n$ movable regions for MAs, as
\begin{subequations}
    \begin{align}
    &\mathbf{C} = \cup \mathbf{C}_{i,j}, \quad i = 1,2,...,N,\quad j = 1,2,...,M, \\
    &\mathbf{t}_{i,j} = (x_i,y_j) \in \mathbf{C}_{i,j},
    \end{align}
\end{subequations}
where
\begin{subequations}
    \begin{align}
    &x_i \in\left[(i-1)\cdot\lfloor N/n\rfloor+1,\ i\cdot\lfloor N/n\rfloor\right], \\
    &y_j \in\left[(j-1)\cdot\lfloor M/m\rfloor+1,\ j\cdot\lfloor M/m\rfloor\right].
    \end{align}
\end{subequations}
The position of each MA is determined as the port where the SINR of the channel is maximized in its local region.

\par For $1\times n$ MAs, we scan the rows of the ports and equally distribute movable regions inside each row for each MA, as
\begin{subequations}
    \begin{align}
    &\mathbf{t}_{u} = (x_u,y),\forall u\in\left[1,n\right]\\
    &x_u \in\left[(u-1)\cdot\lfloor N/n\rfloor+1,\ u\cdot\lfloor N/m\rfloor\right],\\
    &y\in[1,M],
    \end{align}
\end{subequations}
Similarly, for $m\times1$ MAs, movable regions are defined as
\begin{subequations}
    \begin{align}
    &\mathbf{t}_{u} = (x,y_u),\forall u\in\left[1,m\right]\\
    &x\in[1,N],\\
    &y_u \in\left[(u-1)\cdot\lfloor M/m\rfloor+1,\ u\cdot\lfloor M/m\rfloor\right],
    \end{align}
\end{subequations}
from which the best positions are selected with the highest SINR inside movable regions.

\par Fig.~\ref{fig:ma_select_port} shows three examples of two-dimensional MAs, i.e., $2\times2$ planar MAs ($m=n=2$), $2\times1$ linear MAs ($m=2$, $n=1$), and $1\times4$ linear MAs ($m=1$, $n=4$), selected from $32\times32$ candidate ports ($M=N=32$) in the given area.

\subsubsection{Greedy Selection}
\par The greedy selection scheme does not distribute the selectable area into local movable regions for each MA, and directly selects the best $N_t$ channels, in terms of SINR, among all selectable ports in the given area.
In Section~\ref{sec:performance}, we also evaluate the optimal performance of MAs with the greedy selection of antenna positions, as the upper bound. Since the implementation of MA systems barely enables unrestricted movement of all antennas in the given area, the goal of the simulation with the greedy selection is to provide an upper bound for the comparision with the proposed uniform-region SINR-optimized position selection algorithm.

\subsection{Beamforming Algorithm}

\par After the position selection, we perform the beamforming scheme where the Tx is equipped with $N_t$ antennas and one RF chain to communicate with a single-antenna Rx. Denote the channel vector as $\mathbf{h} \in \mathbb{C}^{\mathrm{1 \times N_t}}$, then the received signal of the multi-input-single-output system can be written as
\begin{equation}
\label{eq: system model}
    \mathbf{y} = \mathbf{h}\mathbf{f}_{\rm RF}s + n,
\end{equation}
where $s \in \mathbb{C}$ represents the transmit symbol, and $\mathbf{f}_{\rm RF} \in \mathbb{C}^{N_t \times 1}$ is the analog precoder. $n \sim \mathcal{CN}(0,\sigma_n^2)$ is the additive white Gaussian noise. Denoting the total transmit power as $p$, then the spectral efficiency can be represented as
\begin{equation} \label{eq:SE}
    SE = \log_2(1+\frac{p}{\sigma_n^2}\left|\mathbf{h} \mathbf{f}_{RF}\right|^2).
\end{equation}

\par To maximize the spectral efficiency, the optimal solution of the analog precoder is given by $\mathbf{f}^{\rm opt} = \mathbf{h}/||\mathbf{h}||_2$, where $||\cdot||_2$ is the L2-norm. However, since the analog precoder is implemented by the phase shifters, it should follow the constant modulus constraint, i.e., $|\mathbf{f}_{\mathrm{RF}}(i)| = \frac{1}{\sqrt{N}},\forall i$.
Therefore, the optimization problem becomes
\begin{equation}
\begin{aligned}
    \min_{\mathbf{f}_{\mathrm{RF}}} & ||\mathbf{f}^{\rm opt}-\mathbf{f}_{\mathrm{RF}}||_2^2, \\
    \text { s. t. } & \left|\mathbf{f}_{\mathrm{RF}}(i)\right|=\frac{1}{\sqrt{N}}, \forall i.
\end{aligned}
\end{equation}
The solution is given by~\cite{wu2018hybrid,chen2023hybrid}
\begin{equation}
    \mathbf{f}_{\mathrm{RF}}(i) = \frac{1}{\sqrt{N}} e ^{i \angle{\mathbf{f}^{\text{opt}}(i)}},
\end{equation}
where $\angle{(\cdot)}$ denotes the phase of the element.

\subsection{Evaluation Results} \label{sec:performance}
\begin{figure*}[t]
    \centering
    \subfigure[1x4 MAs.]{
    \includegraphics[width=0.48\linewidth]{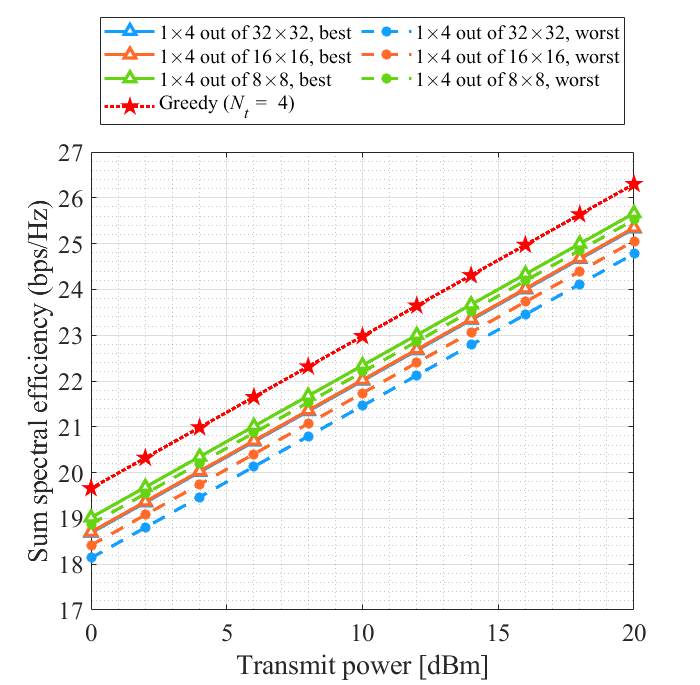}
    }
    \subfigure[4x1 MAs.]{
    \includegraphics[width=0.48\linewidth]{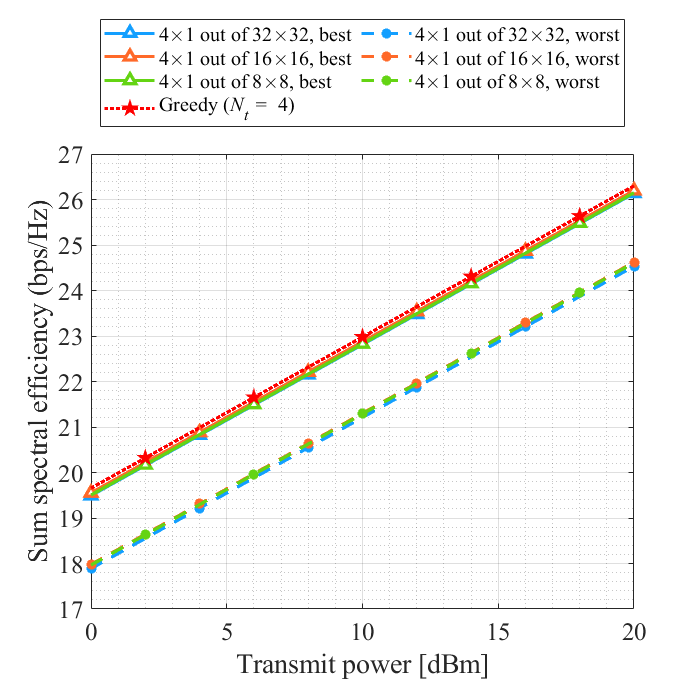}
    }
    \caption{Linear MAs selected from $32\times32$, $16\times16$ and $8\times8$ ports.}
    \label{fig:ULA_out_of_8x8,16x16,32x32}
\end{figure*}

\begin{figure}
    \centering
    \includegraphics[width=\linewidth]{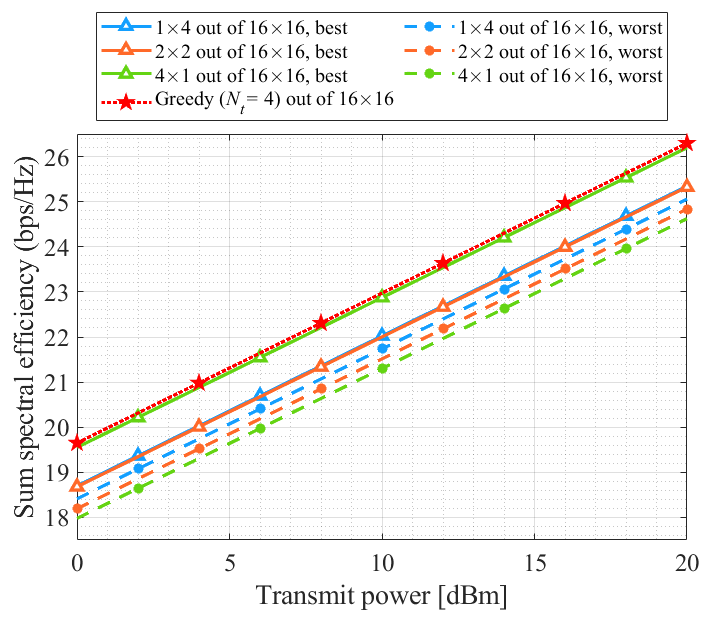}
    \caption{Comparison between MA systems with $N_t=4$.}
    \label{fig:antenna_type_compare}
\end{figure}

\par In this part, we evaluate the performance of linear and planar MA systems in terms of spectral efficiency in~\eqref{eq:SE}. The standard deviation of the noise Gaussian distribution is $\sigma_n^2=4.89\times10^{-6}$. The transmit power varies from 0~dBm to 20~dBm. The discussion is divided into three parts. First, we evaluate and compare the performance of different MA types. Second, the performance achieved by the proposed position selection algorithm is compared with the optimal performance. Third, the MA-enabled improvement of communication performance in terms of spectral efficiency is summarized.

\subsubsection{Comparision among different MA types}
\par First, we examine the performance of $1\times4$ and $4\times1$ linear MAs selected from $32\times32$, $16\times16$ and $8\times8$ ports around the same center.
The result is shown in Fig.~\ref{fig:ULA_out_of_8x8,16x16,32x32} and elaborated as follows. First, since the change of the two-ray channel shows a sequence of power maxima and minima along horizontal ports, the performance of $1\times4$ linear MAs is dependent on the range of selectable ports. Specifically, when candidate ports are more concentrated around the center, where the power is higher at the maxima, the performance of the $1\times4$ MA system is better. For instance, when the transmit power is 0~dBm, the spectral efficiency of the $1\times4$ MA system is 18.6829, 18.7017, and 19.0179~bps/Hz when the selection area is $32\times32$, $16\times16$, and $8\times8$ ports around the center.

\par In contrast, the performance of the $4\times1$ MA system is barely dependent on the range of selectable ports, as the change of channel gain along vertical ports is less significant. Moreover, as long as a column of maxima at the $x$-axis is found, the performance of $4\times1$ linear MAs on this column becomes 99\% close to the optimal performance derived by the greedy selection.

\par Furthermore, we compare the performance of different MA systems under the same $N_t$. To be concrete, $4\times1$, $2\times2$, and $1\times4$ MAs at Tx are simulated. As shown in Fig.~\ref{fig:antenna_type_compare}, the $4\times1$ linear MA system performs the best, which is 99\% close to the optimal performance with the greedy selection of channels. The $2\times2$ planar MA system and the $4\times1$ linear MA system have similar performance that is 95\% of the optimal performance.
To conclude, on account of the measurement scenario, the $n\times1$ linear MA system not only performs better than the $1\times n$ linear MA system, but also approaches the optimal performance under the same number of Tx antennas ($N_t$) in the two-ray wireless channel.

\subsubsection{Comparision with the optimal performance}
\par
\begin{figure}
    \centering
    \includegraphics[width=\linewidth]{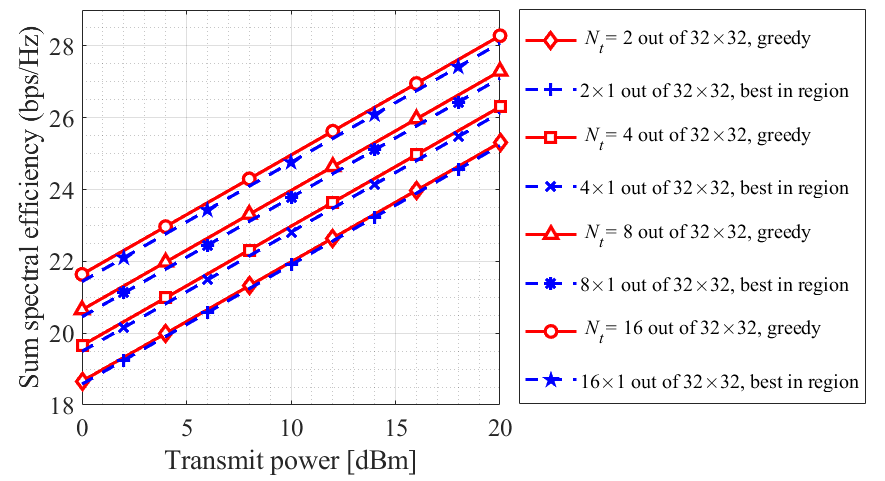}
    \caption{Performance of $n\times 1$ linear MAs with the uniform-region SINR-optimized position selection algorithm, compared with greedy selection.}
    \label{fig:ULA_Nx1_out_of_32x32_greedy}
\end{figure}

 As shown in Fig.~\ref{fig:ULA_Nx1_out_of_32x32_greedy}, the performance of $n\times 1$ linear MAs is compared with the optimal performance, which is derived by greedy selction, i.e., selecting $N_t=n$ best channels out of all candidate channels from $32\times32$ ports to the Rx. The observation is summarized as follows.
\par First, the spectral efficiency increases as $N_t$ increases. Second, the proposed port selection scheme in Section~\ref{sec:port_select}, which equally distributes the given area into movable regions based on the antenna array type, can achieve 99.07\%, 99.05\%, 99.17\%, and 99.57\% of the optimal performance for $2\times1$, $4\times1$, $8\times1$, and $16\times1$ MAs selected from $32\times32$ ports. Hence, the proposed uniform-region position selection scheme not only reduces the complexity by distributing the given area into local regions for each MA, but also is adequate for the $n\times 1$ linear MA system to approximate the optimal performance.

\begin{table}[t]
\caption{Performance improvement for $n\times1$ MAs selected from $N\times N$ candidate ports.}
\label{tab:performance_improvement}
\begin{tabular}{|c|ccccc|}
\hline
\multirow{2}{*}{$n\times1$ MA} & \multicolumn{5}{c|}{$N\times N$ candidate port area}                                                                                                                                                   \\ \cline{2-6} 
                                   & \multicolumn{1}{c|}{\textbf{$32\times32$}} & \multicolumn{1}{c|}{\textbf{$16\times16$}} & \multicolumn{1}{c|}{\textbf{$8\times8$}} & \multicolumn{1}{c|}{\textbf{$4\times4$}} & \textbf{$2\times2$} \\ \hline
\textbf{$2\times1$}                & \multicolumn{1}{c|}{11.48\%}               & \multicolumn{1}{c|}{9.89\%}                & \multicolumn{1}{c|}{9.38\%}              & \multicolumn{1}{c|}{9.19\%}              & 5.79\%              \\ \hline
\textbf{$4\times1$}                & \multicolumn{1}{c|}{8.96\%}                & \multicolumn{1}{c|}{8.75\%}                & \multicolumn{1}{c|}{8.56\%}              & \multicolumn{1}{c|}{8.63\%}              & /                   \\ \hline
\textbf{$8\times1$}                & \multicolumn{1}{c|}{7.72\%}                & \multicolumn{1}{c|}{8.01\%}                & \multicolumn{1}{c|}{8.04\%}              & \multicolumn{1}{c|}{/}                   & /                   \\ \hline
\textbf{$16\times1$}               & \multicolumn{1}{c|}{6.96\%}                & \multicolumn{1}{c|}{7.47\%}                & \multicolumn{1}{c|}{/}                   & \multicolumn{1}{c|}{/}                   & /                   \\ \hline
\end{tabular}
\end{table}

\begin{figure}[t]
    \centering
    \includegraphics[width=\linewidth]{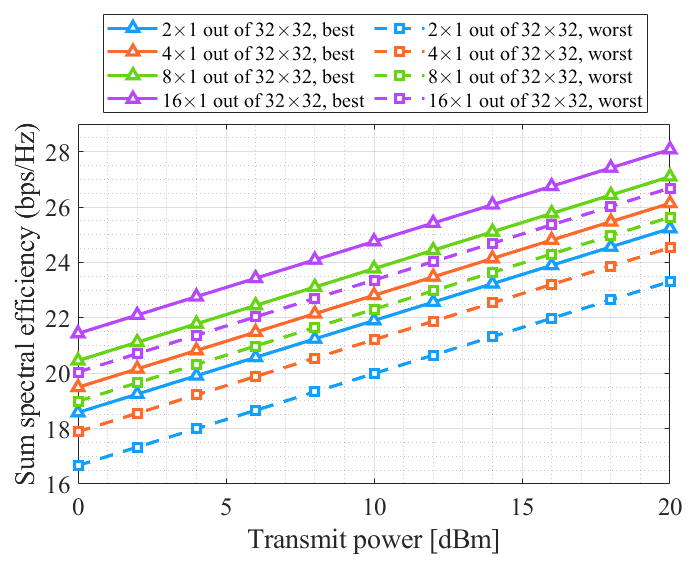}
    \caption{Performance of $n\times 1$ linear MAs across $32\times32$ ports.}
    \label{fig:ULA_Nx1_out_of_32x32}
\end{figure}

\subsubsection{Performance improvement of MA systems}
\par We further summarize the advantage of MA systems over
fixed-position antennas in improving the spectral efficiency by analyzing the improvement of performance with $n\times 1$ linear MAs selected from $N\times N$ candidate ports, compared with the worst case.
The result is summarized in Table~\ref{tab:performance_improvement} and partially shown in Fig.~\ref{fig:ULA_Nx1_out_of_32x32}. On one hand, when the number of antennas decreases in the fixed port area, or the candidate port area expands for the same MA type, the freedom of MA positions and thus the degree of performance improvement increase. For $2\times1$ linear MAs selected from $32\times32$ ports, which maximizes the freedom of choice of antenna positions, SE is increased by 11.48\% at most.
On the other hand, when the number of antennas and the selectable port area both become large, e.g., $n\geq8$ and $N\geq16$, the ports around the center, which have the best channel condition, are allocated to the movable region of one or several single MAs. In this case, the performance improvement is restrained.

\section{Conclusion} \label{section: conclusion}

In this paper, we provided a complete assessment of movable antenna systems in the real-world environment.
First, with a 0.02~mm-resolution two-dimensional movable antenna system, a 60-GHz wideband channel measurement is carried out across $32\times32$ ports at 300~GHz. The multi-path fading, caused by the superposition of LoS and surface-reflected rays, is modeled by the practical two-ray model.
In light of the measurement results, spatial-correlated channel models for the two-dimensional MA system are proposed, statistically parameterized by the complex covariance matrix of measured channels.
Moreover, the low-complexity and near-optimal uniform-region SINR-maximized antenna position selection is proposed. The beamforming algorithm is applied to evaluate the performance of planar and linear MA systems in terms of spectral efficiency.
The results demonstrate the advantage of MAs over fixed-position antennas, by coping with the multi-path fading, in improving channel gains and thus the spectral efficiency by 11.48\%, which reaches 99\% of the optimal performance, across $32\times32$ millimeter-interval ports in the THz wireless channel.



\bibliographystyle{IEEEtran}
\bibliography{bibliography}

\begin{thebibliography}{10}
\providecommand{\url}[1]{#1}
\csname url@samestyle\endcsname
\providecommand{\newblock}{\relax}
\providecommand{\bibinfo}[2]{#2}
\providecommand{\BIBentrySTDinterwordspacing}{\spaceskip=0pt\relax}
\providecommand{\BIBentryALTinterwordstretchfactor}{4}
\providecommand{\BIBentryALTinterwordspacing}{\spaceskip=\fontdimen2\font plus
\BIBentryALTinterwordstretchfactor\fontdimen3\font minus
  \fontdimen4\font\relax}
\providecommand{\BIBforeignlanguage}[2]{{%
\expandafter\ifx\csname l@#1\endcsname\relax
\typeout{** WARNING: IEEEtran.bst: No hyphenation pattern has been}%
\typeout{** loaded for the language `#1'. Using the pattern for}%
\typeout{** the default language instead.}%
\else
\language=\csname l@#1\endcsname
\fi
#2}}
\providecommand{\BIBdecl}{\relax}
\BIBdecl

\bibitem{paulraj2004overview}
A.~Paulraj, D.~Gore, R.~Nabar, and H.~Bolcskei, ``{An overview of MIMO
  communications - a key to gigabit wireless},'' \emph{Proceedings of the
  IEEE}, vol.~92, no.~2, pp. 198--218, Feb. 2004.

\bibitem{heath2016overview}
R.~W. Heath, N.~González-Prelcic, S.~Rangan, W.~Roh, and A.~M. Sayeed, ``{An
  Overview of Signal Processing Techniques for Millimeter Wave MIMO Systems},''
  \emph{IEEE Journal of Selected Topics in Signal Processing}, vol.~10, no.~3,
  pp. 436--453, Apr. 2016.

\bibitem{akyildiz2016realizing}
I.~F. Akyildiz and J.~M. Jornet, ``{Realizing ultra-massive MIMO (1024$\times$
  1024) communication in the (0.06--10) terahertz band},'' \emph{Nano
  Communication Networks}, vol.~8, pp. 46--54, Jun. 2016.

\bibitem{wong2020fluid}
K.-K. Wong, K.-F. Tong, Y.~Zhang, and Z.~Zhongbin, ``{Fluid Antenna System for
  6G: When Bruce Lee Inspires Wireless Communications},'' \emph{Electronics
  Letters}, vol.~56, no.~24, pp. 1288--1290, Nov. 2020.

\bibitem{wong2021fluid}
K.-K. Wong, A.~Shojaeifard, K.-F. Tong, and Y.~Zhang, ``{Fluid Antenna
  Systems},'' \emph{IEEE Transactions on Wireless Communications}, vol.~20,
  no.~3, pp. 1950--1962, Mar. 2021.

\bibitem{wong2022bruce}
K.-K. Wong, K.-F. Tong, Y.~Shen, Y.~Chen, and Y.~Zhang, ``{Bruce Lee-Inspired
  Fluid Antenna System: Six Research Topics and the Potentials for 6G},''
  \emph{Frontiers in Communications and Networks}, vol.~3, Mar. 2022.

\bibitem{wong2022fluid}
K.-K. Wong and K.-F. Tong, ``{Fluid Antenna Multiple Access},'' \emph{IEEE
  Transactions on Wireless Communications}, vol.~21, no.~7, pp. 4801--4815,
  Jul. 2022.

\bibitem{zhu2024movable}
L.~Zhu, W.~Ma, and R.~Zhang, ``{Movable Antennas for Wireless Communication:
  Opportunities and Challenges},'' \emph{IEEE Communications Magazine},
  vol.~62, no.~6, pp. 114--120, Jun. 2024.

\bibitem{zheng2024flexible}
J.~Zheng, J.~Zhang, H.~Du, D.~Niyato, S.~Sun, B.~Ai, and K.~B. Letaief,
  ``{Flexible-Position MIMO for Wireless Communications: Fundamentals,
  Challenges, and Future Directions},'' \emph{IEEE Wireless Communications
  (Early Access)}, pp. 1--9, 2024.

\bibitem{ma2024mimo}
W.~Ma, L.~Zhu, and R.~Zhang, ``{MIMO Capacity Characterization for Movable
  Antenna Systems},'' \emph{IEEE Transactions on Wireless Communications},
  vol.~23, no.~4, pp. 3392--3407, Apr. 2024.

\bibitem{martinez2022towards}
J.~O. Martínez, J.~R. Rodríguez, Y.~Shen, K.-F. Tong, K.-K. Wong, and A.~G.
  Armada, ``{Toward Liquid Reconfigurable Antenna Arrays for Wireless
  Communications},'' \emph{IEEE Communications Magazine}, vol.~60, no.~12, pp.
  145--151, Dec. 2022.

\bibitem{zhuravlev2015experimental}
A.~Zhuravlev, V.~Razevig, S.~Ivashov, A.~Bugaev, and M.~Chizh, ``{Experimental
  simulation of multi-static radar with a pair of separated movable
  antennas},'' \emph{in Proc. of IEEE International Conference on Microwaves,
  Communications, Antennas and Electronic Systems (COMCAS)}, pp. 1--5, Nov.
  2015.

\bibitem{li2022using}
X.~Li, Y.~Zhou, Z.~Shen, B.~Song, and S.~Li, ``{Using a Moving Antenna to
  Improve GNSS/INS Integration Performance Under Low-Dynamic Scenarios},''
  \emph{IEEE Transactions on Intelligent Transportation Systems}, vol.~23,
  no.~10, pp. 17\,717--17\,728, Oct. 2022.

\bibitem{zhu2024modeling}
L.~Zhu, W.~Ma, and R.~Zhang, ``{Modeling and Performance Analysis for Movable
  Antenna Enabled Wireless Communications},'' \emph{IEEE Transactions on
  Wireless Communications}, vol.~23, no.~6, pp. 6234--6250, Jun. 2024.

\bibitem{zhu2024multiuser}
L.~Zhu, W.~Ma, B.~Ning, and R.~Zhang, ``{Movable-Antenna Enhanced Multiuser
  Communication via Antenna Position Optimization},'' \emph{IEEE Transactions
  on Wireless Communications}, vol.~23, no.~7, pp. 7214--7229, Jul. 2024.

\bibitem{hu2024fluid}
G.~Hu, Q.~Wu, K.~Xu, J.~Ouyang, J.~Si, Y.~Cai, and N.~Al-Dhahir, ``{Fluid
  Antennas-Enabled Multiuser Uplink: A Low-Complexity Gradient Descent for
  Total Transmit Power Minimization},'' \emph{IEEE Communications Letters},
  vol.~28, no.~3, pp. 602--606, Mar. 2024.

\bibitem{pi2023multiuser}
X.~Pi, L.~Zhu, Z.~Xiao, and R.~Zhang, ``{Multiuser Communications with
  Movable-Antenna Base Station Via Antenna Position Optimization},'' \emph{in
  Proc. of IEEE Globecom Workshops (GC Wkshps)}, pp. 1386--1391, Dec. 2023.

\bibitem{chai2022port}
Z.~Chai, K.-K. Wong, K.-F. Tong, Y.~Chen, and Y.~Zhang, ``{Port Selection for
  Fluid Antenna Systems},'' \emph{IEEE Communications Letters}, vol.~26, no.~5,
  pp. 1180--1184, May 2022.

\bibitem{ye2024fluid}
Y.~Ye, L.~You, J.~Wang, H.~Xu, K.-K. Wong, and X.~Gao, ``{Fluid
  Antenna-Assisted MIMO Transmission Exploiting Statistical CSI},'' \emph{IEEE
  Communications Letters}, vol.~28, no.~1, pp. 223--227, Jan. 2024.

\bibitem{mei2024movable}
W.~Mei, X.~Wei, B.~Ning, Z.~Chen, and R.~Zhang, ``{Movable-Antenna Position
  Optimization: A Graph-Based Approach},'' \emph{IEEE Wireless Communications
  Letters}, vol.~13, no.~7, pp. 1853--1857, Jul. 2024.

\bibitem{zhu2023movable}
L.~Zhu, W.~Ma, and R.~Zhang, ``{Movable-Antenna Array Enhanced Beamforming:
  Achieving Full Array Gain With Null Steering},'' \emph{IEEE Communications
  Letters}, vol.~27, no.~12, pp. 3340--3344, Dec. 2023.

\bibitem{ma2024multi}
W.~Ma, L.~Zhu, and R.~Zhang, ``{Multi-Beam Forming With Movable-Antenna
  Array},'' \emph{IEEE Communications Letters}, vol.~28, no.~3, pp. 697--701,
  Mar. 2024.

\bibitem{kang2024deep}
J.-M. Kang, ``{Deep Learning Enabled Multicast Beamforming With Movable Antenna
  Array},'' \emph{IEEE Wireless Communications Letters}, vol.~13, no.~7, pp.
  1848--1852, Jul. 2024.

\bibitem{chen2023joint}
X.~Chen, B.~Feng, Y.~Wu, D.~W. Kwan~Ng, and R.~Schober, ``{Joint Beamforming
  and Antenna Movement Design for Moveable Antenna Systems Based on Statistical
  CSI},'' \emph{in Proc. of IEEE Global Communications Conference}, pp.
  4387--4392, Dec. 2023.

\bibitem{qin2024antenna}
H.~Qin, W.~Chen, Z.~Li, Q.~Wu, N.~Cheng, and F.~Chen, ``{Antenna Positioning
  and Beamforming Design for Fluid Antenna-Assisted Multi-User Downlink
  Communications},'' \emph{IEEE Wireless Communications Letters}, vol.~13,
  no.~4, pp. 1073--1077, Apr. 2024.

\bibitem{hu2024movable}
G.~Hu, Q.~Wu, J.~Ouyang, K.~Xu, Y.~Cai, and N.~Al-Dhahir,
  ``{Movable-Antenna-Array-Enabled Communications With CoMP Reception},''
  \emph{IEEE Communications Letters}, vol.~28, no.~4, pp. 947--951, Apr. 2024.

\bibitem{wang2024far}
Y.~Wang, S.~Sun, and C.~Han, ``{Far- and Near-Field Channel Measurements and
  Characterization in the Terahertz Band Using a Virtual Antenna Array},''
  \emph{IEEE Communications Letters}, vol.~28, no.~5, pp. 1186--1190, May 2024.

\bibitem{wang2022thz}
Y.~Wang, Y.~Li, Y.~Chen, Z.~Yu, and C.~Han, ``{0.3 THz Channel Measurement and
  Analysis in an L-shaped Indoor Hallway},'' \emph{in Proc. of IEEE ICC}, pp.
  1--6, Jun. 2022.

\bibitem{li2022channel}
Y.~Li, Y.~Wang, Y.~Chen, Z.~Yu, and C.~Han, ``{Channel Measurement and Analysis
  in an Indoor Corridor Scenario at 300 GHz},'' \emph{in Proc. of IEEE ICC},
  pp. 1--6, Jun. 2022.

\bibitem{wu2018hybrid}
X.~Wu, D.~Liu, and F.~Yin, ``{Hybrid Beamforming for Multi-User Massive MIMO
  Systems},'' \emph{IEEE Transactions on Communications}, vol.~66, no.~9, pp.
  3879--3891, Sep. 2018.

\bibitem{chen2023hybrid}
Y.~Chen, R.~Li, C.~Han, S.~Sun, and M.~Tao, ``{Hybrid Spherical- and
  Planar-Wave Channel Modeling and Estimation for Terahertz Integrated UM-MIMO
  and IRS Systems},'' \emph{IEEE Transactions on Wireless Communications},
  vol.~22, no.~12, pp. 9746--9761, Dec. 2023.

\end{thebibliography}

\vfill

\end{document}